\documentclass[journal]{IEEEtran}
\usepackage[T1]{fontenc}
\usepackage[utf8]{inputenc}
\usepackage{graphicx}
\usepackage[hyphens]{url}
\usepackage[caption=false,font=footnotesize,hangindent=10pt]{subfig}
\usepackage{amsmath,amsfonts,amsthm,amssymb}
\usepackage{mathtools}
\usepackage[linesnumbered,vlined,boxed,ruled]{algorithm2e}
\usepackage[colorlinks]{hyperref}
\usepackage{color,comment,xparse}
\usepackage{tabu}
\usepackage{array}
\usepackage{tikz}
\usetikzlibrary{shapes.arrows,shapes.geometric,positioning,fit,calc,patterns}

\graphicspath{{./}{figure/}}
\def\Snospace~{\S{}}

\SetCommentSty{myAlgComment}
\SetKwProg{Fn}{Function}{:}{}
\SetAlFnt{\small}
\SetAlCapFnt{\small}
\SetAlCapNameFnt{\small}

\NewDocumentCommand{\twofigs}{o m m m}{%
  \subfloat[#2\IfValueTF{#1}{\label{#1_1}}{}]{\includegraphics[width=.5\linewidth]{#3}} &
  \subfloat[#2\IfValueTF{#1}{\label{#1_2}}{}]{\includegraphics[width=.5\linewidth]{#4}}}
\newcommand{\header}[1]{\smallskip\noindent\textbf{#1}}

\DeclarePairedDelimiter{\floor}{\lfloor}{\rfloor}

\newcommand{\E}{\mathbb{E}}
\newcommand{\indr}[1]{\mathbf{1}(#1)}
\newcommand{\CG}{\mathcal{G}}
\newcommand{\Guu}{\CG_\text{uu}}
\newcommand{\Guc}{\CG_\text{uc}}
\newcommand{\Euu}{\mathcal{E}_\text{uu}}
\newcommand{\Euc}{\mathcal{E}_\text{uc}}
\newcommand{\CL}{\mathcal{L}}
\newcommand{\pt}{p_\triangle}
\newcommand{\BBin}{\text{BetaBin}}

\newtheorem{theorem}{Theorem}

\begin{document}
\title{Tracking Triadic Cardinality Distributions for Burst Detection in High-Speed
  Multigraph Streams}

\author{Junzhou Zhao, Pinghui Wang, John C.S. Lui, Don Towsley, and Xiaohong Guan}

\markboth{Journal of \LaTeX\ Class Files,~Vol.~6, No.~1, January~2007}%
{Shell \MakeLowercase{\textit{et al.}}: Bare Demo of IEEEtran.cls for Journals}

\maketitle

\begin{abstract}
In everyday life, we often observe unusually frequent interactions among people
before or during important events, e.g., people send/receive more greetings to/from
their friends on holidays than regular days.
We also observe that some videos or hashtags suddenly go viral through people's
sharing on online social networks (OSNs).
Do these seemingly different phenomena share a common structure?
All these phenomena are associated with sudden surges of user interactions in
networks, which we call ``{\em bursts}'' in this work.
We uncover that the emergence of a burst is accompanied with the formation of
triangles in some properly defined networks.
This finding motivates us to propose a new and robust method to detect bursts on
OSNs.
We first introduce a new measure, ``{\em triadic cardinality distribution}'',
corresponding to the fractions of nodes with different numbers of triangles, i.e.,
triadic cardinalities, in a network.
We show that this distribution not only changes when a burst occurs, but it also
has a robustness property that it is immunized against common spamming social-bot
attacks.
Hence, by tracking triadic cardinality distributions, we can more reliably detect
bursts than simply counting interactions on OSNs.
To avoid handling massive activity data generated by OSN users during the triadic
tracking, we design an efficient ``{\em sample-estimate}'' framework to provide
maximum likelihood estimate on the triadic cardinality distribution.
We propose several sampling methods, and provide insights about their performance
difference, through both theoretical analysis and empirical experiments on real
world networks.

\end{abstract}

\begin{IEEEkeywords}
burst detection,
sampling methods,
data streaming algorithms
\end{IEEEkeywords}

\IEEEpeerreviewmaketitle

\section{Introduction}
\label{sec:intro}

Online social networks (OSNs) have become ubiquitous platforms that provide various
ways for users to interact over the Internet, such as tweeting tweets, sharing
links, messaging friends, commenting on posts, and mentioning/replying to other
users (i.e., @someone).
When intense user interactions take place in a short time period, there will be a
surge in the volume of user activities in an OSN.
Such a surge of user activity, which we call a ``{\em burst}'' in this work,
usually relates to emergent events that are occurring or about to occur in the real
world.
For example, Michael Jackson's death on June 25, 2009 triggered a global outpouring
of grief on Twitter~\cite{Harvey2009}, and the event even crashed Twitter for
several minutes~\cite{Shiels2009}.
In addition to bursts caused by real world events, some bursts arising from OSNs
can also cause enormous social impact in the real world.
For example, the 2011 England riots, in which people used OSNs to organize,
resulted in $3,443$ crimes across London due to this disorder~\cite{LondonRiots}.
Hence, detecting bursts in OSNs is an important task, both for OSN managers to
monitor the operation status of an OSN, as well as for government agencies to
anticipate any emergent social disorder.

Typically, there are two types of user interactions in OSNs.
First is the interaction between users (we refer to this as {\em user-user
  interaction}), e.g., a user sends a message to another user, while the second is
the interaction between a user and a media content piece (we refer to this as {\em
  user-content interaction}), e.g., a user (re-)posts a video link.
Examples of bursts caused by these two types of interactions include, many
greetings being sent/received among people on Christmas Day, and videos suddenly
becoming viral after one day of sharing in an OSN.
At first sight, detecting such bursts in an OSN is not difficult.
For example, a naive way to detect bursts caused by user-user interactions is to
{\em count} the number of pairwise user interactions within a time window, and
report a burst if the volume lies above a given threshold.
However, this method is vulnerable to spamming social-bot attacks~\cite{Chu2010,
  Grier2010,Stringhini2010a,Boshmaf2011,Thomas2011,Beutel2013a}, which can suddenly
generate a huge amount of spamming interactions in the OSN.
Hence, this method can result in many {\em false alarms} due to the existence of
social bots.
Similar problem also exists when detecting bursts caused by user-content
interactions.
Many previous works on burst detection are based on idealized
assumptions~\cite{Kleinberg2002,Yi2005,Parikh2008,Eftekhar2013} and simply ignore
the existence of social bots.

\header{Present work.}
The primary goal of this work is to leverage a special {\em triangle structure},
which is a feature of human interaction and behavior, to design a robust burst
detection method that is immune against common social-bot attacks.
We first describe the triangle structure shared by both types of user interactions.

\begin{figure}[t]
  \centering
  \tikzstyle{usr}=[draw, thick, circle, minimum size=8pt, inner sep=0pt]
\tikzstyle{txt}=[inner sep=1pt, anchor=west, align=center]
\tikzstyle{arr}=[->, >=latex]

\subfloat[normal\label{f:ib1}]{
  \begin{tikzpicture}[every node/.style={font=\scriptsize}]
    \path[use as bounding box] (0,0) rectangle (2.2, 2.3);

    \begin{scope}[xshift=3mm,yshift=5mm]
      \node(a) [usr] {$a$};
      \node(b) [usr, above = 1.2cm of a] {$b$};
      \node(c) [usr, above right = .7cm and .3cm of a] {$u$};
      \node(d) [usr, above right = .1cm and .7cm of c] {$d$};
      \node(e) [usr, below = .7cm of d] {$e$};
      \draw[blue,thick] (a) -- (b) -- (c) -- (d) -- (e) -- (c)-- (a);
      \node[txt, above right= 0 and .3 of b] (au) {a user};
      \node[txt, below right= 0 and .4 of a.center] (ai) {an interaction};
      \draw[arr] (au.west) -- (b);
      \draw[arr] (ai.170) -- (.2,.3);
    \end{scope}

  \end{tikzpicture}
}
\subfloat[interaction burst\label{f:ib2}]{
  \begin{tikzpicture}[every node/.style={font=\scriptsize}]
    \path[use as bounding box] (0,0) rectangle (2.2, 2.3);

    \begin{scope}[xshift=3mm,yshift=5mm]
      \node(a) [usr] {};
      \node(b) [usr, above = 1.2cm of a] {};
      \node(c) [usr, above right = .7cm and .3cm of a] {};
      \node(d) [usr, above right = .1cm and .7cm of c] {};
      \node(e) [usr, below = .7cm of d] {};
      \draw[red,thick] (a) -- (b) -- (c) -- (a) -- (e) -- (c) -- (d) -- (e);
      \draw[red,thick] (b) -- (d) -- (a);
    \end{scope}
  \end{tikzpicture}
}
\subfloat[triadic cardinality dist.\label{f:ib3}]{\includegraphics[width=.4\linewidth]{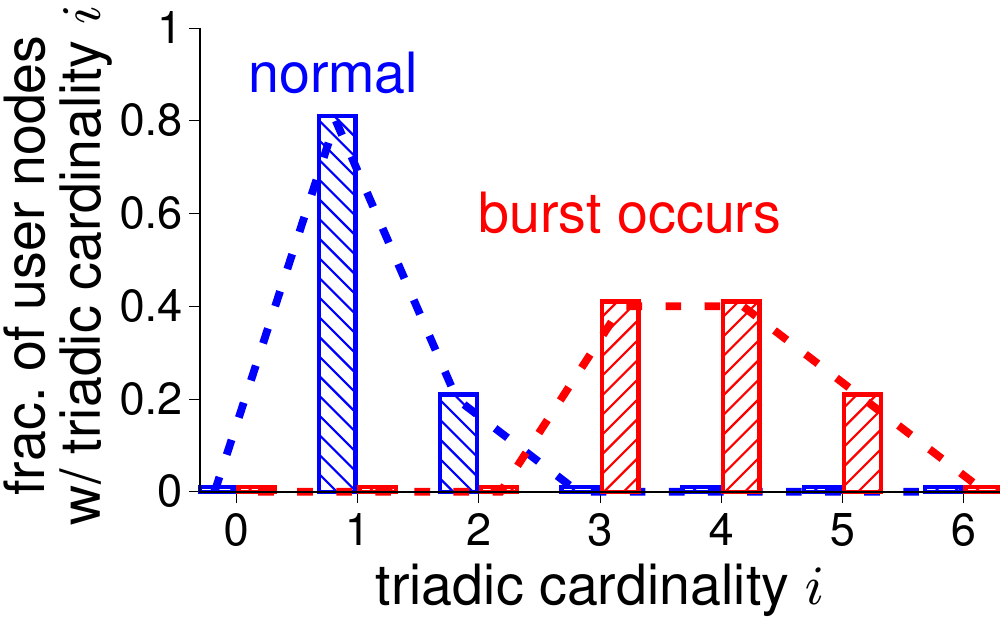}}

  \caption{Interaction burst and interaction triangle, in which edges in (a) and
    (b) represent interactions among users.}
  \label{fig:ib_exam}
\end{figure}
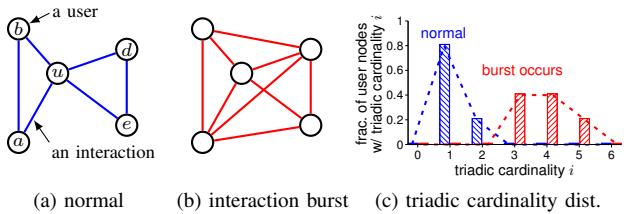

\header{Interaction triangles in user-user interactions.}
Humans form social networks with larger clustering coefficients than those in
random networks~\cite{Watts1998} because social networks exhibit many \emph{triadic
  closures}~\cite{Kossinets2006}.
This is due to the social phenomenon of ``{\em friends of my friends are also my
  friends}''.
Since user-user interactions usually take place along social links, this property
implies that user-user interactions should also exhibit many triadic closures
(which we will verify in later experiments).
In other words, when a group of users suddenly become active, or we say an {\em
  interaction burst} occurs, in addition to observing the rise of volume of
pairwise interactions, we expect to also observe many interactions among three
neighboring users, i.e., many {\em interaction triangles} form if we consider an
edge of an interaction triangle to be a user-user interaction.
This is illustrated in Fig.~\ref{f:ib1} when no interaction burst occurs, while in
Fig.~\ref{f:ib2}, an interaction burst occurs.
In contrast, activities generated by social bots do not possess many triangles
since social bots typically select their targets randomly from an
OSN~\cite{Boshmaf2011,Thomas2011}.

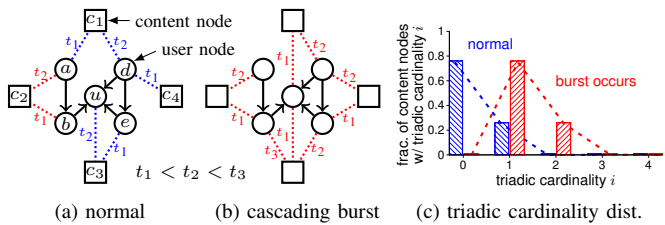
\begin{figure}[t]
  \centering
  \tikzstyle{usr}=[draw, thick, circle, minimum size=8pt, inner sep=0pt]
\tikzstyle{csd}=[draw, thick, rectangle, minimum size=8pt, inner sep=0]
\tikzstyle{csdlne}=[densely dotted, thick]
\tikzstyle{lb}=[midway, inner sep=#1, font=\tiny]
\tikzset{lb/.default=1pt}
\tikzstyle{txt}=[inner sep=2pt, anchor=west, align=center]
\tikzstyle{arr}=[->,>=latex]

\newcommand{\tm}[1]{$t_{\textsc{#1}}$}

\subfloat[normal\label{f:cb1}]{%
  \begin{tikzpicture}[every node/.style={font=\scriptsize}]

    \path[use as bounding box] (0,0) rectangle (2.5,2.5);

    \begin{scope}[xshift=1.25cm,yshift=1.25cm]
    \node(f) [usr] {$u$};
    \node(a) [usr, above left = 4pt and 5pt of f] {$a$};
    \node(b) [usr, below left = 4pt and 5pt of f] {$b$};
    \node(d) [usr, above right = 4pt and 5pt of f] {$d$};
    \node(e) [usr, below right = 4pt and 5pt of f] {$e$};

    \draw[->,thick] (a) -- (b);
    \draw[->,thick] (b) -- (f);
    \draw[->,thick] (d) -- (e);
    \draw[->,thick] (e) -- (f);
    \draw[->,thick] (d) -- (f);

    \node(c1) [csd, above = 7mm of f] {$c_1$};
    \node(c2) [csd, left = 7mm of f] {$c_2$};
    \node(c3) [csd, below = 7mm of f] {$c_3$};
    \node(c4) [csd, right = 7mm of f] {$c_4$};

    \node[txt,right=3mm of c3] {$t_1<t_2<t_3$};

    \draw[blue,csdlne](a)--(c1) node[lb=0,left,pos=.6]{\tm{1}};
    \draw[blue,csdlne](d)--(c1) node[lb,right,pos=.6]{\tm{2}};
    \draw[red,csdlne](a)--(c2) node[lb,above,pos=.6]{\tm{2}};
    \draw[red,csdlne](b)--(c2) node[lb,below,pos=.6]{\tm{1}};
    \draw[blue,csdlne](f)--(c3) node[lb=0,left]{\tm{2}};
    \draw[blue,csdlne](e)--(c3) node[lb,right,pos=.6]{\tm{1}};
    \draw[blue,csdlne](d)--(c4) node[lb,above,pos=.7]{\tm{1}};

    \node[txt,right=.3 of c1] (ac) {content node};
    \node[txt, above right=0 and .3 of d] (au) {user node};
    \draw[arr] (ac.west) -- (c1.east);
    \draw[arr] (au.183) -- (d.north east);
    \end{scope}

  \end{tikzpicture}
}%
\subfloat[cascading burst\label{f:cb2}]{%
  \begin{tikzpicture}[every node/.style={font=\scriptsize}]
    \path[use as bounding box] (0,0) rectangle (2.5,2.5);

    \begin{scope}[xshift=1.25cm,yshift=1.25cm]
    \node(f) [usr] {};
    \node(a) [usr, above left = 4pt and 5pt of f] {};
    \node(b) [usr, below left = 4pt and 5pt of f] {};
    \node(d) [usr, above right = 4pt and 5pt of f] {};
    \node(e) [usr, below right = 4pt and 5pt of f] {};
    \draw[->,thick] (a) -- (b);
    \draw[->,thick] (b) -- (f);
    \draw[->,thick] (d) -- (e);
    \draw[->,thick] (e) -- (f);
    \draw[->,thick] (d) -- (f);
    \node(c1) [csd, above = 7mm of f] {};
    \node(c2) [csd, left = 7mm of f] {};
    \node(c3) [csd, below = 7mm of f] {};
    \node(c4) [csd, right = 7mm of f] {};

    \draw[red,csdlne](f)--(c1) node[lb=0,left]{\tm{1}};
    \draw[red,csdlne](d)--(c1) node[lb,right]{\tm{2}};
    \draw[red,csdlne](a)--(c2) node[lb,above,pos=.6]{\tm{2}};
    \draw[red,csdlne](b)--(c2) node[lb,below,pos=.6]{\tm{1}};
    \draw[red,csdlne](b)--(c3) node[lb=0,left,pos=.7]{\tm{3}};
    \draw[red,csdlne](f)--(c3) node[lb=0,left]{\tm{1}};
    \draw[red,csdlne](e)--(c3) node[lb,right,pos=.6]{\tm{2}};
    \draw[red,csdlne](d)--(c4) node[lb,above,pos=.7]{\tm{2}};
    \draw[red,csdlne](e)--(c4) node[lb=2pt,below,pos=.7]{\tm{1}};
    \end{scope}
  \end{tikzpicture}
}%
\subfloat[triadic cardinality dist.\label{f:cb3}]{\includegraphics[width=.4\linewidth]{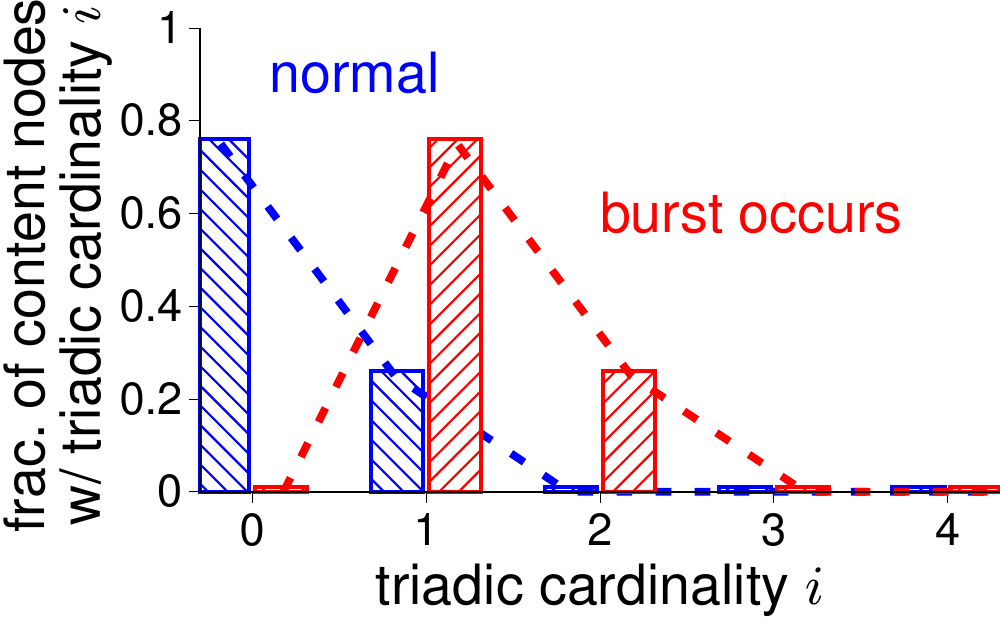}}

  \caption{Cascading burst and influence triangle.}
  \label{fig:cb_exam}
\end{figure}

\header{Influence triangles in user-content interactions.}
Similar triangle structure can also be observed in bursts caused by user-content
interactions.
We say that a media content piece becomes \emph{bursty} if many users interact with
it in a short time period.
There are many reasons why a user interacts with a piece of media content.
Here, we are particularly interested in the case where one user {\em influences}
another user to interact with the content, a.k.a., the cascading
diffusion~\cite{Leskovec2007b} or word-of-mouth spreading~\cite{Rodrigues2011}.
It is known that many emerging news stories arising from OSNs are related to this
mechanism such as the story about the killing of Osama bin
Laden~\cite{Tsotsis2011}.
We find that a bursty media content piece formed by this mechanism is associated
with triangle formations in a network.
To illustrate this, consider Fig.~\ref{f:cb1}, in which there are five user nodes
$\{a,b,d,e,u\}$ and four content nodes $\{c_1,c_2,c_3,c_4\}$.
A directed edge between two users means that one follows another, and an undirected
edge labeled with a timestamp between a user node and a content node represents an
interaction between the user and the content at the labeled time.
We say content node $c$ has an {\em influence triangle} if there exist two users
$a,b$ such that $a$ follows $b$ and $a$ interacts with $c$ {\em later} than $b$
does.
In other words, the reason $a$ interacts with $c$ is due to the influence of $b$ on
$a$.
In Fig.~\ref{f:cb1}, only $c_2$ has an influence triangle, the others have no
influence triangle, meaning that the majority of user-content interactions are not
due to influence; while in Fig.~\ref{f:cb2}, every content node is part of at least
one influence triangle, meaning that many content pieces are spreading in a
cascading manner in the OSN.
From the perspective of an OSN manager who wants to know the operation status of
the OSN, if the OSN suddenly switches to a state similar to Fig.~\ref{f:cb2} (from
a previous state similar to Fig.~\ref{f:cb1}), he knows that a {\em cascading
  burst} is present on the network.

\header{Characterizing bursts.}
So far, we find a common structure shared by different types of bursts: the
emergence of {\em interaction bursts} (caused by user-user interaction) and {\em
  cascading bursts} (caused by user-content interaction) are both accompanied with
the formation of triangles, i.e., interaction or influence triangles, in
appropriately defined networks.
This finding motivates us to characterize patterns of bursts in an OSN by
characterizing the triangle statistics of a network, which we called the {\em
  triadic cardinality distribution}.

{\em Triadic cardinality} of a node in a network, e.g., a user node in
Fig.~\ref{f:ib1} or a content node in Fig.~\ref{f:cb1}, is the number of triangles
that it belongs to.
The triadic cardinality distribution then characterizes the fractions of nodes with
certain triadic cardinalities.
When a burst occurs, because many new interaction/influence triangles are formed,
we will observe that some nodes' triadic cardinalities increase, and this results
in the distribution ``shifting'' to right, as illustrated in Figs.~\ref{f:ib3}
and~\ref{f:cb3}.
The triadic cardinality distribution provides succinct summary information to
characterize burst patterns of a large scale OSN.
Hence, by tracking triadic cardinality distributions, we can detect the presence of
bursts.

In this paper, we assume that user interactions are aggregated chronologically to
form a {\em social activity stream}, which can be considered as an abstraction of a
{\em tweet stream} in Twitter, or a {\em news feed} in Facebook.
We aim to calculate triadic cardinality distributions from this stream.
The challenge is that when a network is large or users are active, the social
activity stream will be of high speed.
For example, the speed of the Twitter's tweets stream can be as high as $5,700$
tweets per second on average, $143,199$ tweets per second during the peak time, and
about $500$ million to $12$ billion tweets are aggregated per
day~\cite{Krikorian2013}.
To handle such a high-speed social activity stream, we design a sample-estimate
framework, which can provide {\em maximum likelihood estimates} of the triadic
cardinality distribution using sampled data.
Our system works in a near-real-time fashion, and is demonstrated to be accurate
and efficient.

\header{Contributions.}
In this work, we make the following contributions:
\begin{itemize}
\item We find a useful and robust measure, triadic cardinality distribution, that
  can be used to characterize burst patterns of user interactions in a large OSN.
  It has a robustness property that is immunized against common spamming social-bot
  attacks.
\item We design a unified sample-estimate framework that is able to provide maximum
  likelihood estimates of the triadic cardinality distribution.
  Under this framework, we study two types of stream sampling methods, and provide
  insights about their performance difference, through calculating the Fisher
  information matrices, and empirical evaluations.
\item We conduct extensive experiments using real world data to demonstrate the
  usefulness of triadic cardinality distribution, and prove the effectiveness of
  our sample-estimate framework.
\end{itemize}

The remainder of the work will proceed as follows.
In~\autoref{sec:problem}, we formally define the triadic cardinality distribution,
and give an overview about our sample-estimate solution.
In~\autoref{sec:sampling}, we design two types of stream sampling methods to reduce
storage complexity and improve computational efficiency.
We then elaborate a maximum likelihood estimation method in~\autoref{sec:estimate},
and obtain the Cram\'{e}r-Rao lower bound in~\autoref{sec:crlb}.
We provide detailed validations of our methods in~\autoref{sec:experiment},
including a real world application on detecting bursts during the 2014 Hong Kong
occupy central movement.
Finally, \autoref{sec:relatedwork} summarizes some related work,
and~\autoref{sec:conclusion} concludes.

\section{Problem Formulation}
\label{sec:problem}

We first formally define the notions of social activity stream and triadic
cardinality distribution mentioned in Introduction.
Then we give an overview of our sample-estimate framework.

\subsection{Social Activity Stream}

We represent an OSN by a simple graph $G=(V,E,C)$, where $V$ is a set of users, $E$
is a set of relations among users, and $C$ is a set of media content such as
hashtags and video links.
Here, a relation between two users can be undirected like the friend relationship
in Facebook, or directed like the follower relationship in Twitter.

Users in the OSN generate {\em social activities}, e.g., interact with other users
in $V$, or content in $C$.
We denote a social activity by $a\in V\times(V\cup C)\times [0,\infty)$.
Here {\em user-user interaction}, $a=(u,v,t)$, corresponds to user $u$ interacting
with user $v$ at time $t$; and {\em user-content interaction}, $a=(u,c,t)$,
corresponds to user $u$ interacting with content $c$ at time $t$.
These social activities are then aggregated chronologically to form a {\em social
  activity stream}, denoted by $S=\{a_1,a_2,\ldots\}$, where $a_l$ denotes the
$l$-th social activity in the stream.

\subsection{Triadic Cardinality Distribution}

We introduce two {\em interaction multigraphs} formed by the two types of user
interactions respectively.
Triadic cardinality distribution is then defined on these two interaction
multigraphs.

\header{Interaction multigraphs.}
Within a time window (e.g., an hour, a day, or a week), user-user interactions in
stream $S$ form a multigraph $\Guu=(V,\Euu)$, where $V$ is the original set of
users, and $\Euu$ is a multiset consisting of user-user interactions in the window.
The triadic cardinality of a user $u\in V$ is the number of interaction triangles
related to $u$ in $\Guu$.
For example, user $u$ in Fig.~\ref{f:ib1} has triadic cardinality two, and all
other users have triadic cardinality one.

Likewise, user-content interactions also form a multigraph $\Guc=(V\cup C,E\cup
\Euc)$ in a time window.
Unlike $\Guu$, the node set in $\Guc$ includes both user nodes $V$ and content
nodes $C$, and the edge set includes user relations $E$ and a multiset $\Euc$
denoting user-content interactions in the window.
Note that in $\Guc$, triadic cardinality is only defined for content nodes, and the
triadic cardinality of a content node $c\in C$ is the number of influence triangles
related to $c$ in $\Guc$.
For example, in Fig.~\ref{f:cb1}, content $c_2$ has triadic cardinality one, and
all other content nodes have triadic cardinality zero.

\header{Triadic cardinality distribution.}
Let $\theta = (\theta_0,\ldots,\theta_W)$ and $\vartheta = (\vartheta_0,\ldots,
\vartheta_{W'})$ denote the triadic cardinality distributions on $\Guu$ and $\Guc$
respectively.
Here, $\theta_i$ (or $\vartheta_i$) is the fraction of user (or content) nodes with
triadic cardinality $i$ in $\Guu$ (or $\Guc$), and $W$ (or $W'$) denotes the
maximum triadic cardinality.

The importance of the triadic cardinality distribution lies in its capability of
providing succinct summary information to characterize burst patterns in a large
scale OSN as we mentioned previously.
By tracking triadic cardinality distributions, we will discover burst occurrences
in an OSN.

\subsection{Overview of Our Sample-Estimate Framework}

\begin{figure}[htp]
  \centering
  \begin{tikzpicture}[
  every node/.style={thick, font=\footnotesize},
  txt/.style={align=center, inner sep=2pt},
  blk/.style={draw, rectangle, minimum height=6mm, minimum width=12mm,
    text depth=.25ex},
  arr/.style={draw, thin, single arrow, single arrow head extend=#1, fill=gray!30,
    align=center, minimum height=8mm}]

  \node(osn) {\includegraphics[width=8mm]{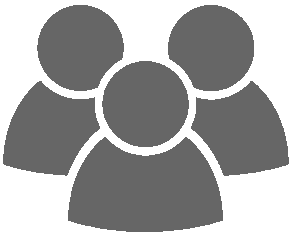}};
  \node[arr=1.5ex,right =0 of osn] (sa) {social\\activities};
  \node[blk,right =0 of sa] (sp) {sample};
  \node[arr=1ex,single arrow head extend=1ex,right =0 of sp] (a1) {};
  \node[blk,right =0 of a1] (es) {estimate};
  \node[arr=1ex,right =0 of es] (a2) {};
  \node[draw, dashed, fit={(sp) (es)}, inner sep=5pt] (gp) {};

  \node[txt, left=5pt of sa.east] {$\CG$};
  \node[txt, left=3pt of a1.east] {$\CG'$};

  \node[txt, below=8pt of osn] (osn_txt) {OSN};
  \node[txt] at (osn_txt -| gp) {measurement};

  \begin{scope}[shift={($(a2.east)+(.05,-.4)$)}]
    \draw[<->,thick] (0,.9)--(0,0)--(.9,0);
    \draw[very thick] (.05,.7) .. controls (.2,.2) .. (.7,.05);
    \node[txt] at (.5,.5) {$\hat\theta,\hat\vartheta$};
    \coordinate (A) at (.4,-1.2em) {};
  \end{scope}
  \node[txt] at (osn_txt -| A) {triadic cardinality \\ distribution};
\end{tikzpicture}

  \caption{A sample-estimate framework}
  \label{fig:overview}
\end{figure}
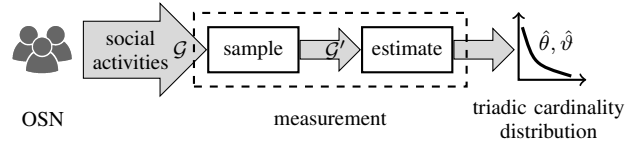

We propose an online solution capable of tracking the triadic cardinality
distribution from a high-speed social activity stream in a near-real-time fashion,
as illustrated in Fig.~\ref{fig:overview}.

Our system consists of two stages.
In the first stage, we sample a social activity stream in a time window maintaining
only summary statistics, and in the second stage, we construct an estimate of the
triadic cardinality distribution from the summary statistics at the end of a time
window.
The advantage of this approach is that it reduces the amount of data need to be
stored and processed in the system, and enables detecting bursts in a
near-real-time fashion.

\section{Stream Sampling Methods}
\label{sec:sampling}

In this section, we elaborate the sampling module in our system, and design two
types of stream sampling methods.
The purpose of sampling is to reduce the computational cost in handling the massive
amount of data in a high-speed stream.

\subsection{Identical Triangle Sampling (ITS)}

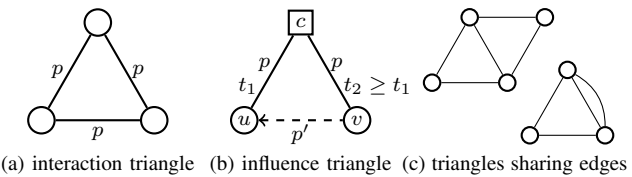
\begin{figure}[htp]
  \centering \footnotesize
  \tikzstyle{usr}=[draw,thick,circle,minimum size=10pt,inner sep=0]
\tikzstyle{susr}=[usr,minimum size=6.2pt]
\tikzstyle{con}=[draw,thick,rectangle,minimum size=9pt,inner sep=0]
\tikzstyle{lb}=[midway,inner sep=2pt]
\tikzstyle{txt}=[align=center, inner sep=2pt]

\subfloat[interaction triangle\label{f:t1_pd}]{
  \begin{tikzpicture}
    \path[use as bounding box] (-.5,-.3) rectangle (2,1.5);
    \node(u) [usr] at (0,0) {};
    \node(v) [usr] at (1.5,0) {};
    \node(w) [usr] at (.75,1.3) {};
    \draw[thick] (u) -- (v) node[lb,below] {$p$};
    \draw[thick] (u) -- (w) node[lb,left] {$p$};
    \draw[thick] (v) -- (w) node[lb,right] {$p$};
  \end{tikzpicture}
}%
\subfloat[influence triangle\label{f:t2_pd}]{
  \begin{tikzpicture}
    \path[use as bounding box] (-.5,-.3) rectangle (2,1.5);

    \node[usr] (u) at (0,0) {$u$};
    \node[usr] (v) at (1.5,0) {$v$};
    \node[con] (c) at (.75,1.3) {$c$};
    \draw[<-,dashed,thick] (u) -- (v) node[lb,below,inner sep=1pt] {$p'$};
    \draw[thick] (u)--(c) node[lb,left,pos=.6] {$p$} node[lb,left,pos=.3] {$t_1$};
    \draw[thick] (v)--(c) node[lb,right,pos=.6] {$p$}
      node[lb,right,pos=.3]{$t_2\geq t_1$};
  \end{tikzpicture}
}%
\subfloat[triangles sharing edges\label{f:t_dependence}]{
  \begin{tikzpicture}

    \path[use as bounding box] (-.3,-.3) rectangle (2.5,1.5);

    \begin{scope}[yshift=5mm]
      \node(a) [susr] at (0,0) {};
      \node(b) [susr] at (1,0) {};
      \node(c) [susr] at (.5,.87) {};
      \node(d) [susr] at (1.5,.87) {};
      \draw (a) -- (b) -- (c) -- (a);
      \draw (b) -- (d) -- (c);
    \end{scope}

    \begin{scope}[xshift=1.3cm,yshift=-2mm]
      \node(a) [susr] at (0,0) {};
      \node(b) [susr] at (1,0) {};
      \node(c) [susr] at (.5,.87) {};
      \draw (a) -- (b) -- (c) -- (a);
      \draw (b) to [bend right] (c);
    \end{scope}
  \end{tikzpicture}
}

  \caption{Sampling triangles.
    A solid edge represents an interaction in $\Euu\cup\Euc$, and a dashed edge
    represents a social edge in $E$.}
  \label{fig:triangle}
\end{figure}

The simplest stream sampling method should work as follows.
We toss a biased coin for each coming social activity $a\in S$.
We keep $a$ with probability $p$, and ignore it with probability $1-p$.
Hence, each social activity is independently sampled, and at the end of the time
window, only a fraction $p$ of the stream is kept.
When social activities are sampled, triangles in graphs $\Guu$ and $\Guc$ are
sampled accordingly.
Obviously, an interaction triangle is sampled with identical probability $p^3$, as
illustrated in Fig.~\ref{f:t1_pd}.

For influence triangle, we need a few more considerations.
First, an influence triangle consists of two user-content interaction edges in
$\Euc$ and one social edge in $E$.
Second, stream sampling only applies to edges in $\Euc\cup \Euu$, and edges in $E$
are not sampled because they do not appear in the stream.
In Fig.~\ref{f:t2_pd}, suppose we have sampled two user-content interactions
$(u,c,t_1)$ and $(v,c,t_2)$, and assume $t_1\leq t_2$, i.e., user $u$ interacts
with $c$ earlier than $v$.
To determine whether content $c$ has an influence triangle formed by users $u$ and
$v$, we need to check whether (directed) edge $(v,u)$ exists in $E$.
This can be done by querying neighbors of one of the two users in the OSN.
For example, in Twitter, we query \emph{followees} of $v$ and check whether $v$
follows $u$; or in Facebook, we query friends of $v$ and check whether $u$ is a
friend of $v$.
Sometimes this query cost is expensive if we do not own $G$ and need to call the
API provided by the OSN.
To reduce this query cost, we check a user pair with probability $p'$.
This is equivalent to sampling a social edge in $E$ with probability $p'$,
conditioned on the two associated user-content interactions having been sampled.
Thus, an influence triangle is sampled with identical probability $p^2p'$.
In summary, we have the following result.

\begin{theorem}\label{th:pd}
  If we independently sample each social activity in stream $S$ with probability
  $p$, and check a user relation in $E$ with probability $p'$, then each
  interaction (influence) triangle in $\Guu$ ($\Guc$) is sampled with identical
  probability
  \begin{equation}
    \pt^\text{ITS}=
    \begin{cases}
      p^3   & \text{for an interaction triangle},\\
      p^2p' & \text{for an influence triangle}.
    \end{cases}
  \end{equation}
\end{theorem}

\header{A more efficient ITS method.}
An obvious drawback of the previous sampling method is that an interaction triangle
is sampled with probability cubic to the edge sampling probability.
This means that an interaction triangle is hardly sampled if $p$ is too small.
In fact, we can increase triangle sampling probability to $p^2$ (and still keep
each edge being sampled with probability $p$) by leveraging a clever {\em colorful
  triangle sampling} method~\cite{Pagh2012}.
Let $N=1/p$ be an integer, and $[N]$ represents a set of $N$ colors.
Define a hash function $h\colon V\mapsto [N]$ that maps a node to one of these $N$
colors uniformly at random.
During sampling, for a coming social activity $a=(u,v,t)$, we keep $a$ if
$h(u)=h(v)$, and drop $a$ otherwise.
We can see that a user-user interaction is still sampled with probability $p$, but
an interaction triangle is now sampled with probability $p^2$, and hence it is more
efficient in collecting triangles from edge samples.
For influence triangle, we can let $p'=1$, i.e., we check every sampled user pair
(similar to gSH~\cite{Ahmed2014}), and an influence triangle is also sampled with
probability $p^2$.
We will refer to ITS method with colorful triangle sampling as ITS-color in the
following discussion.

\begin{remark}
  In both ITS and ITS-color, although triangles are sampled identically, they may
  {\em not} be sampled {\em independently}, such as the cases two triangles share
  edges in Fig.~\ref{f:t_dependence}.
  We will consider this issue in detail later.
\end{remark}

\subsection{Harvesting Triangles by Subgraph Sampling (SGS)}

The ITS based methods are easy to implement, and they are already used for counting
the triangles in a large network~\cite{Tsourakakis2009,Pagh2012}.
However, ITS based methods have drawbacks when they are used for estimating the
triadic cardinality distribution.
One main drawback is that, because ITS samples each triangle with identical
probability, the sampling will be biased towards nodes belonging to many triangles.
That is, nodes with larger triadic cardinalities are more likely to be sampled, and
for nodes with small triadic cardinalities, the triangles these nodes belonging to
will be seldom sampled.
This hence may incur large estimation error for nodes with small triadic
cardinalities.
To address this weakness, we propose another triangle sampling method that
leverages interaction multigraphs and social graph in a different way, which we
call the subgraph sampling (SGS) method.

For interaction triangle, assume that we are only interested in user-user
interactions along social edges.
Then SGS works as follows.
At the beginning of a time window, we first sample a set of user samples that each
user is sampled with probability $p_n$ (and this step can be independently
implemented on social graph $G$ using well-studied graph sampling
techniques~\cite{Gjoka2011}).
For each sampled user, a subgraph induced by the user and the user's neighbors in
$G$ is maintained, i.e., each edge in the induced subgraph is an social edge in
$G$.
During stream sampling, for each social activity $a=(u,v,t)$, if $(u,v)$ is an edge
in one of these subgraphs, we keep $a$; otherwise $a$ is dropped.
In this way, interaction triangles related to the user sample are kept completely.

Similar procedure is also applied to sample influence triangles, and here we aim to
keep complete influence triangles related to each sampled content node.
To sample a content node with probability $p_n$, we need to store content nodes
seen so far in a Bloom filter for ease of testing whether a coming content node is
new or not.
For a coming social activity $a=(u,c,t)$, if $c$ is already marked as a sampled
content node, we keep $a$; if $c$ is new (i.e., $c$ is not found in the Bloom
filter), we mark $c$ as a content node sample with probability $p_n$ and save $a$,
otherwise we drop $a$.
No matter $a$ is saved or discarded, if $c$ is new, we need to store $c$ in the
Bloom filter.
So, we can see that social activities related to sampled content nodes are all
kept, if we also check the existence of social edges with probability $p'=1$, then
influence triangles related to sampled content nodes are kept completely.

SGS method thus keeps every triangle a node sample belonging to, whatever the node
sample has large or small triadic cardinality.
We will later develop rigorous method to compare the performance of ITS based
methods and SGS method.

\subsection{Statistics of Sampled Data}

Both ITS (including ITS-color) and SGS can be thought of as sampling edges in
multigraphs $\Guu$, $\Guc$, and social graph $G$, in different manners.
In ITS, interaction edges $e\in\Euu\cup\Euc$ are independently sampled with
probability $p$, and social edges $e'\in E$ are sampled with conditional
probability $p'$.\footnote{Conditioned on the two user-content interaction edge
  being sampled before checking the social edge.}
In SGS, we first sample a collection of user/content nodes with probability $p_n$,
and then only keep triangles related to these sampled user/content nodes.

At the end of the time window, we obtain two sampled multigraphs $\Guu'$ and
$\Guc'$.
Calculating the triadic cardinalities for nodes in these smaller graphs is much
efficient than on the original graphs.
For the sampled graph $\Guu'$, we calculate triadic cardinality for each (sampled)
user node in ITS (SGS), and obtain statistics $g=(g_0,\ldots,g_W)$, where $g_j$,
$0\leq j\leq W$, denotes the number of user nodes belonging to $j$ triangles in
graph $\Guu'$.
Similar statistics are also obtained from $\Guc'$, denoted by
$f=(f_0,\ldots,f_{W'})$ (where $f_j$ is the number of content nodes belonging to
$j$ influence triangles in graph $\Guc'$).
We only need to store $g$ and $f$ in main memory and use them to estimate $\theta$
and $\vartheta$ in the next section.

\section{Maximum Likelihood Estimate}
\label{sec:estimate}

In this section, we elaborate the estimation module in our system, and derive a
maximum likelihood estimate (MLE) of the triadic cardinality distribution using
statistics obtained in the sampling step.
The estimation in this section can be viewed as an analog of the network flow size
distribution
estimation~\cite{Duffield2003,Ribeiro2006,Tune2011,Wang2014a,Veitch2015}, in which
a packet in a flow is viewed to be a triangle a node belonging to.
However, in our case, triangle samples are not independent, and a node may have no
triangles at all.
These issues complicate the estimation design, and we will describe how to solve
these issues in this section.

Note that we only discuss how to estimate $\theta$ using $g$, as the MLE of
$\vartheta$ using $f$ is easily obtained using a similar approach.
To estimate $\theta$, we first consider the easier case where graph size $|V|=n$ is
known.
Later, we extend our analysis to the case where $|V|$ is unknown.

\subsection{A General MLE Framework when Graph Size is Known}

Recall that $g_j$, $0\leq j\leq W$, is the number of nodes having $j$ sampled
triangles in graph $\Guu'$.
First, note that observing a node with $j$ sampled triangles in graph $\Guu'$
implies that the node has at least $j$ triangles in graph $\Guu$.
Second, we also need to pay special attention to $g_0$, which is the number of
nodes with no triangle observed in graph $\Guu'$.
Due to sampling, some nodes may be unobserved (e.g., no edge attached to the node
is sampled in ITS, or the node is not sampled in SGS), and these ``evaporated''
nodes are actually ``observed'' to have zero triangle because graph $\Guu$ has $n$
nodes.
Hence, we need to include these evaporated nodes in $g_0$, i.e.,
\[
  g_0 \coloneqq g_0 + n-\sum_{j=0}^W g_j = n-\sum_{j=1}^W g_j.
\]

To derive a MLE of $\theta$, we use a conditional probability to model the sampling
process.
For a randomly chosen node, let $X$ denote the number of triangles to which it
belongs in $\Guu$, and let $Y$ denote the number of triangles observed during
sampling.
Let $b_{ji}\triangleq P(Y=j|X=i)$ for $0\leq j\leq i$ be the conditional
probability that a node has $j$ sampled triangles in $\Guu'$ given that it has $i$
triangles in the original graph $\Guu$.
Then the probability of observing a node to have $j$ sampled triangles is
\begin{equation}\label{eq:pY}
  P(Y=j) \!=\! \sum_{i=j}^W P(Y=j|X=i)P(X=i) \!=\! \sum_{i=j}^W b_{ji}\theta_i.
\end{equation}
Then, the log-likelihood of observations $\{Y_l=y_l\}_{k=1}^n$, where $Y_l=y_l$
denotes the $k$-th node having $y_l$ sampled triangles, yields
\begin{equation}\label{eq:likelihood}
  \CL(\theta)\triangleq\log P(\{Y_l=y_l\}_{k=1}^n)
  =\sum_{j=0}^W g_j \log\sum_{i=j}^W b_{ji}\theta_i.
\end{equation}

The MLE of $\theta$ can then be obtained by maximizing $\CL(\theta)$ with respect
to $\theta$ under the constraint that $\sum_{i=0}^W\theta_i = 1$.
To solve the likelihood maximization problem, we face two challenges: (1) What are
the specific formulas of $b_{ji}$ for the sampling methods we have proposed
previously?
(2) Note that it is impossible to obtain a closed form solution maximizing
Eq.~\eqref{eq:likelihood}, so how should we design an algorithm to maximize
Eq.~\eqref{eq:likelihood} conveniently?

\subsubsection{Sampling Model Specification}

We specify the sampling models $b_{ji}$ for ITS and SGS, respectively.
We start with SGS for its simplicity.

\header{SGS:} Remember that SGS keeps all the triangles of each sampled node, and a
node is sampled with probability $p_n$.
Hence, if we observe a node belonging to $j>0$ triangles in $\Guu'$, the node must
have $i=j$ triangles in $\Guu$, and this occurs with probability $p_n$.
If we observe a node having no triangle in $\Guu'$, i.e., $j=0$, there are two
possibilities, i.e., the node indeed has no triangle in $\Guu$, $i=0$, or the node
is not sampled.
Therefore,
\[
  b_{ji} = \begin{cases}
    1,     & \text{if }j=i=0,         \\
    1-p_n, & \text{if }j=0\wedge i>0, \\
    p_n,   & \text{if }j=i,           \\
    0,     & \text{otherwise}.
    \end{cases}
\]

\header{ITS and ITS-color:} In ITS or ITS-color, each triangle is sampled with
identical probability, denoted by $\pt$.
Sampling a triangle can be thought of as a Bernoulli trial with success probability
$\pt$.
If Bernoulli trials are independent, meaning that triangles are independently
sampled, then $b_{ji}=P(Y=j|X=i)$ should follow the standard binomial distribution
parameterized by $i$ and $\pt$.
Unfortunately, independence does not hold for triangles sharing edges, as
illustrated in Fig.~\ref{f:t_dependence}.
As a result, it is non-trivial to derive an accurate sampling model $b_{ji}$ for
ITS and ITS-color due to the dependence among triangle samples.
To deal with this issue, we propose to approximate the {\em sums of dependent
  Bernoulli random variables} by a Beta-binomial
distribution~\cite{Beta-bin,Yu2002}, which is given by\footnote{Strictly, we
  should use the notation $\BBin(j|i, \pt/\alpha, (1-\pt)/\alpha)$ according
  to~\cite{Beta-bin}, however, we simplify it to $\BBin(j|i, \pt, \alpha)$.
}
\begin{align*}
  &\hspace{-.2in}\BBin(j|i, \pt, \alpha) \\
  &\triangleq\binom{i}{j}
    \frac{\prod_{s=0}^{j-1}(s\alpha + \pt)\prod_{s=0}^{i-j-1}(s\alpha + 1 -\pt)}
    {\prod_{s=0}^{i-1}(s\alpha + 1)}
\end{align*}
and $\prod_0^{-1}\triangleq 1$.
The above distribution parameterized by $\alpha\geq 0$ allows pairwise identically
distributed Bernoulli random variables to have covariance $\alpha
\pt(1-\pt)/(1+\alpha)$.
It reduces to the standard binomial distribution when $\alpha=0$.
Hence, for ITS and ITS-color, the sampling model is approximated by
\[
  b_{ji}(\alpha) = \BBin(j|i,\pt,\alpha), \quad j\in [0,i].
\]

\subsubsection{MLE via EM Algorithm}

To solve the second challenge, we propose to use the expectation-maximization (EM)
algorithm to obtain the MLE in a more convenient way.
For general consideration, we use $b_{ji}(\alpha)$ to denote the sampling model.

If we already know that the $k$-th node has $x_l$ triangles in $\Guu$, i.e.,
$X_l=x_l$, then the complete likelihood of observations $\{(Y_l,X_l)\}_{l=1}^n$ is
\begin{align*}
  &P(\{(Y_l,X_l)\}_{l=1}^n) = \prod_{l=1}^n P(Y_l=y_l, X_l=x_l) \\
  &= \prod_{j=0}^W \prod_{i=j}^W P(Y=j, X=i)^{z_{ij}}
    = \prod_{j=0}^W \prod_{i=j}^W \left[b_{ji}(\alpha)\theta_i\right]^{z_{ij}}
\end{align*}
where $z_{ij}=\sum_{l=1}^n \indr{x_l=i\wedge y_l=j}$ is the number of nodes with
$i$ triangles and $j$ of them being sampled; $\indr{\cdot}$ is the indicator
function.
The complete log-likelihood is
\begin{equation}\label{eq:c_log-likelihood}
  \CL_c(\theta,\alpha) \triangleq
  \sum_{j=0}^W \sum_{i=j}^W z_{ij}\log\left[ b_{ji}(\alpha)\theta_i \right].
\end{equation}
Here, we can treat $\{X_l\}_{l=1}^n$ as hidden variables, and apply the EM
algorithm to calculate the MLE.

\header{E-step.}
We calculate the expectation of the complete log-likelihood in
Eq.~\eqref{eq:c_log-likelihood} with respect to hidden variables $\{X_l\}_l$,
conditioned on data $\{Y_l\}_l$ and previous estimates $\theta^{(t)}$ and
$\alpha^{(t)}$.
That is
\[
  Q(\theta,\alpha;\theta^{(t)},\alpha^{(t)}) \triangleq
  \sum_{j=0}^W \sum_{i=j}^W
  \E[z_{ij}|\theta^{(t)},\alpha^{(t)}] \log\left[ b_{ji}(\alpha)\theta_i \right].
\]
Here, $\E[z_{ij}|\theta^{(t)},\alpha^{(t)}]$ can be viewed as the average number of
nodes that have $i$ triangles in $\Guu$, of which $j$ are sampled.
Because
\begin{align*}
  & \hspace{-0.2in} P(X=i|Y=j,\theta^{(t)},\alpha^{(t)}) \\
  &=\frac{P(Y=j|X=i,\alpha^{(t)})P(X=i|\theta^{(t)})}
     {\sum_{i'}P(Y=j|X=i',\alpha^{(t)})P(X=i'|\theta^{(t)})} \\
  &=\frac{b_{ji}(\alpha^{(t)})\theta_i^{(t)}}
     {\sum_{i'} b_{ji'}(\alpha^{(t)})\theta_{i'}^{(t)}}
     \triangleq p_{i|j}
\end{align*}
and we have observed $g_j$ nodes belonging to $j$ sampled triangles, then
$\E[z_{ij}|\theta^{(t)},\alpha^{(t)}] = g_jp_{i|j}$.

\header{M-step.}
We now maximize $Q(\theta,\alpha;\theta^{(t)},\alpha^{(t)})$ with respect to
$\theta$ and $\alpha$ subject to the constraint $\sum_{i=0}^W \theta_i=1$.
After the $\log$ operation, $\theta$ and $\alpha$ are well separated.
Hence, we obtain
\begin{align*}
  \theta_i^{(t+1)}
  &= \arg\max_\theta Q(\theta,\alpha;\theta^{(t)},\alpha^{(t)}) \\
  &= \frac{\sum_{j=0}^i \E[z_{ij}|\theta^{(t)},\alpha^{(t)}]}
    {\sum_{j=0}^W \sum_{i'=j}^W \E[z_{i'j}|\theta^{(t)},\alpha^{(t)}]},
    \quad 0\leq i\leq W,
\end{align*}
and $\alpha^{(t+1)}=\arg\max_\alpha Q(\theta,\alpha;\theta^{(t)},\alpha^{(t)})$ can
be solved using classic gradient ascent methods.

Alternating iterations of the E-step and M-step, EM algorithm converges to a
solution, which is a local maximum of $\CL(\theta,\alpha)$.
We denote this solution by $\hat\theta$ and $\hat\alpha$.

\subsection{MLE when Graph Size is Unknown}
\label{subsec:mle_n_unknown}

When the graph size is unknown, one can use probabilistic counting methods such as
loglog counting~\cite{Durand2003} to obtain an estimate of graph size from the
stream, and then apply our previously developed method to obtain estimate
$\hat\theta$.
Note that this introduces additional statistical errors to $\hat\theta$ due to the
inaccurate estimate of the graph size.
In what follows, we slightly reformulate the problem and develop a method that can
simultaneously estimate both the graph size and the triadic cardinality
distribution from the sampled data.

When the graph size is unknown, we cannot calibrate $g_0$ because ``evaporated''
nodes are indeed unobservable in this case.
There is no clear relationship between an unsampled node and its triadic
cardinality.
As a result, we cannot easily model the absence of nodes by $\theta$.
If we observe a node having no triangle after sampling, we cannot reason out which
way caused the observation, the node has no triangle in the original graph, or the
triangles the node belonging to are not sampled.
This difficulty hence complicates the estimation design.

To solve this issue, we need to slightly reformulate our problem: (1) Instead of
estimating the total number of nodes in $\Guu$, we estimate the number of nodes
belonging to at least one triangle in $\Guu$, denoted by $n_+$; (2) We estimate the
triadic cardinality distribution $\theta^+=(\theta_1^+,\ldots,\theta_W^+)$, where
$\theta_i^+$ is the fraction of nodes with $i$ triangles over the nodes having at
least one triangle in $\Guu$.

\header{Estimating $n_+$.}
Under the Beta-binomial model, the probability that a node has $i$ triangles in
$\Guu$, of which none are sampled, is
\[
  q_i(\alpha) \triangleq P(Y=0|X=i) = b_{0i}(\alpha).
\]
Then, the probability that a node has triangles in $\Guu$, of which none are
sampled, is
\[
  q(\theta^+,\alpha)\triangleq P(Y=0|X\geq 1) = \sum_{i=1}^Wq_i(\alpha)\theta_i^+.
\]
Because there are $\sum_{j=1}^W g_j$ nodes having been observed to have at least
one sampled triangle, $n_+$ can be estimated by
\begin{equation}\label{eq:nplus}
  \hat{n}_+ = \frac{\sum_{j=1}^W g_j}{1-q(\theta^+,\alpha)}.
\end{equation}

Note that estimator~\eqref{eq:nplus} relies on $\theta^+$ and $\alpha$, and we can
estimate them using the following procedures.

\header{Estimating $\theta^+$ and $\alpha$.}
We discard $g_0$ and only use $g^+\triangleq (g_1,\ldots,g_W)$ to estimate
$\theta^+$ and $\alpha$.
The basic idea is to derive the likelihood for nodes that are observed to have at
least one sampled triangle, i.e., $\{Y_l=y_l\colon y_l\geq 1\}$.
In this case, the probability that a node has $X=i$ triangles, and $Y=j$ of them
are sampled, conditioned on $Y\geq 1$, is
\begin{align*}
  P(Y=j|X=i,Y\geq 1) &= \frac{P(Y=j|X=i)}{P(Y\geq 1|X=i)} \\
                     &= \frac{b_{ji}(\alpha)}{1-q_i(\alpha)}
                       \triangleq a_{ji}(\alpha), \quad j\geq 1.
\end{align*}
Then the probability that a node is observed to have $j$ sampled triangles,
conditioned on $Y\geq 1$, is
\begin{align*}
  &\hspace{-0.2in} P(Y=j|Y\geq 1) \\
  &=\sum_{i=j}^W P(Y=j|X=i, Y\geq 1)P(X=i|Y\geq 1)\\
  &=\sum_{i=j}^W a_{ji}(\alpha)\phi_i,
\end{align*}
where
\begin{equation}\label{eq:phi}
  \phi_i \triangleq P(X=i|Y\geq 1)
  = \frac{\theta_i^+[1 - q_i(\alpha)]}
  {\sum_{i''=1}^W \theta_{i'}^+[1 - q_{i'}(\alpha)]},
\end{equation}
is the distribution of observed node triadic cardinalities.
Now it is straightforward to obtain the previously mentioned likelihood.
Furthermore, we can leverage our previously developed EM algorithm by replacing
$\theta_i$ by $\phi_i$, $b_{ji}$ by $a_{ji}$, to obtain MLEs for $\phi$ and
$\alpha$.
We omit these details, and directly provide the final EM iterations:
\[
  \phi_i^{(t+1)}
  = \frac{\sum_{j=1}^i \E[z_{ij}|\phi^{(t)},\alpha^{(t)}]}
  {\sum_{j=1}^W \sum_{i'=j}^W \E[z_{i'j}|\phi^{(t)},\alpha^{(t)}]},
  \quad i\geq 1,
\]
where
\[
  \E[z_{ij}|\phi^{(t)},\alpha^{(t)}]
  = \frac{g_ja_{ji}(\alpha^{(t)})\phi_i^{(t)}}
  {\sum_{i'=j}^W a_{ji'}(\alpha^{(t)})\phi_{i'}^{(t)}},
  \quad i\geq j\geq 1,
\]
and $\alpha^{(t+1)}=\arg\max_\alpha Q(\phi,\alpha;\phi^{(t)},\alpha^{(t)})$ is
solved using gradient ascent methods.

Once EM converges, we obtain estimates $\hat\phi$ and $\hat\alpha$.
The estimate for $\theta^+$ is then obtained by Eq.~\eqref{eq:phi}, i.e.,
\begin{equation}\label{eq:rescaling}
  \hat\theta_i^+ =
  \frac{\hat\phi_i/[1 - q_i(\hat\alpha)]}
  {\sum_{i'=1}^W\hat\phi_{i'}/[1 - q_{i'}(\hat\alpha)]},
  \quad 1\leq i\leq W.
\end{equation}
Finally, $\hat{n}_+$ is obtained by the estimator in Eq.~\eqref{eq:nplus}.

\subsection{Logarithmic Binning Simplification}

In our previous study~\cite{Zhao2015c}, we have observed that triadic cardinality
distributions in many real-world networks exhibit heavy tails.
Thus, it is better to characterize them in the logarithmic scale.
That is, instead of estimating the fraction of nodes having exact triadic
cardinality $i$, we may want to estimate the fraction of nodes with triadic
cardinality in $\log_2$ scaled bins.
We aim to estimate the fraction of nodes having triadic cardinality in the $k$-th
bin $[2^k,2^{k+1})$, denoted by $\theta^+_k$, for $k = 0,1,\ldots,K$ where
$K\triangleq\floor{\log_2W}$.
If we allow $i=0$ as in the case where graph size is known, we define the first bin
to be $\{0\}$ assigning to bin $k=-1$, and use $\theta_k, k=-1,0,\ldots,K$, to
represent the binned triadic cardinality distribution.

In the logarithmic binning simplification, for each $i$ in the $k$-th bin, we
assume that $\theta_i$ has the same value, and $\theta_i = 2^{-k}\theta_k$ for
$k\geq 0$.
We further define
\begin{align*}
  b_{jk}
  &\triangleq P(Y=j|X\in\text{bin}(k)) \\
  &= \sum_{i=2^k}^{2^{k+1}-1}P(Y=j|X=i)P(X=i|X\in\text{bin}(k)) \\
  &= 2^{-k}\sum_{i=2^k}^{2^{k+1}-1} b_{ji}
\end{align*}
for $k\geq 0$.
For $k=-1$, we define $b_{jk}=1$ if $j=0$ and $0$ otherwise.
Similar to Eq.~\eqref{eq:pY}, the probability of observing a node having $j$
triangles after sampling becomes $P(Y=j) = \sum_{k=-1}^K b_{jk}\theta_k$.
Thus, it is straightforward to obtain a MLE of $\theta_k$ using previously
developed methods.
Similar analysis can also be applied to estimate $\theta_k^+$, and hence is
omitted.

The logarithmic binning simplification reduces the number of parameters from $W+1$
to $\floor{\log_2W}+1$ that allows us to consider large triadic cardinality bound
$W$ in large networks.
Meanwhile, $\{\theta_k\}$ and $\{\theta_k^+\}$ are actually smoothed versions of
$\{\theta_i\}$ and $\{\theta_i^+\}$, which we will observe in experiments.

\section{Asymptotic Estimation Error Analysis}
\label{sec:crlb}

To evaluate the performance of MLEs using different sampling methods, this section
devotes to analyze the asymptotic estimation error of the MLEs by calculating the
Cram\'{e}r-Rao lower bound (CRLB) of $\hat\theta$ and $\hat\theta^+$.
It is well-known that MLE is asymptotically Gaussian centered at the true value
with variance the CRLB, and the Cram\'{e}r-Rao theorem further states that the mean
squared error of any unbiased estimator is lower bounded by the CRLB, which is the
inverse of the Fisher information (see~\cite[Chapter 2]{Trees2001} for more
details).

Intuitively, the Fisher information can be thought of as the amount of information
that observations $\{Y_l\}$ carry about unobservable parameters $\theta$ (or
$\theta^+$) upon which the probability distribution of the observations depends.
When graph size is known, the Fisher information of observations $\{Y_l\}$ is a
$(W+1)\times (W+1)$ square matrix $J(\theta)$ whose $ir$-th element is given by
\[
  J_{ir}(\theta)
  \triangleq\E_Y \left[
    \frac{\partial\log P(\{Y_l\}_l|\theta)}{\partial\theta_i}
    \frac{\partial\log P(\{Y_l\}_l|\theta)}{\partial\theta_r}
  \right].
\]
In our problem, $J_{ir}(\theta)$ can be further simplified to
\begin{align*}
  J_{ir}(\theta)
  &=n\sum_{j=0}^W
    \frac{\partial\log P(Y|\theta)}{\partial\theta_i}
    \frac{\partial\log P(Y|\theta)}{\partial\theta_r}P(Y=j) \\
  &=n\sum_{j=0}^W\frac{b_{ji}(\alpha)b_{jr}(\alpha)}{P(Y=j)}.
\end{align*}
When graph size is unknown, the Fisher information matrix $J(\theta^+)$ is a
$W\times W$ matrix.
To obtain $J(\theta^+)$, we can first obtain the Fisher information matrix
regarding to $\phi$, denoted by $J(\phi)$, using an approach similar to above (by
replacing $n$ by $n^+$).
Then $J(\phi)$ and $J(\theta^+)$ are known to have the following
relationship~\cite{crlb}
\[
  J(\theta^+)^{-1} = \nabla H J(\phi)^{-1} \nabla H^T,
\]
where $\nabla H$ is the Jacobian matrix, and its $ir$-th element is given by
$\partial\theta_i^+(\phi)/\partial\phi_r$ (and $\theta_i^+(\phi)$ is given by
Eq.~\eqref{eq:rescaling}).

The inverse constrained Fisher information of $\theta$ with constraint
$\sum_i\theta_i=1$ is then obtained by
\[
  I(\theta) = J(\theta)^{-1} - \theta\theta^T,
\]
where the term $\theta\theta^T$ corresponds to the accuracy gain due to constraint
$\sum_i\theta_i=1$ (see~\cite{Gorman1990,Tune2011} for more details).
Then, mean squared error of an estimator $\hat\theta$ is lower bounded by the
diagonal elements of $I(\theta)$, i.e., $\E[(\hat\theta_i - \theta_i)^2] \geq
I_{ii}(\theta)$.
Similar relation also holds for $\hat\theta^+$.

MLE is asymptotically efficient, and CRLB is its asymptotic variance.
We are thus able to leverage CRLB to compare the asymptotic estimation accuracy of
MLEs using different sampling methods.

\section{Experiments and Validations}
\label{sec:experiment}

In this section, we first empirically verify the claims we have made previously.
Then, we validate the proposed estimation methods on several real-world networks.
Finally, we illustratively show the usefulness of triadic cardinality distribution
in detecting bursts during the Hong Kong Occupy Central movement in Twitter.

\subsection{Analyzing Bursts in Enron Dataset}
\label{ssec:enron}

In the first experiment, we use a public email communication dataset to empirically
show how bursts in networks can change the triadic cardinality distribution, and
verify our claims previously made.

\subsubsection{\textbf{Enron email dataset}}
The Enron email dataset~\cite{Klimt2004} includes the entire email communications
(e.g., who sent an email to whom at what time) of the Enron corporation from its
startup to bankruptcy.
The used dataset is carefully cleaned by removing spamming accounts/emails and
emails with incorrect timestamps.
The cleaned dataset contains $22,477$ email accounts and $164,081$ email
communications between Jan 2001 and Apr 2002.
We use this dataset to study patterns of bursts caused by email communications
among people, i.e., by user-user interactions.

\subsubsection{\textbf{Observations from data}}
\begin{figure}[htp]
\centering
\includegraphics[width=.9\linewidth]{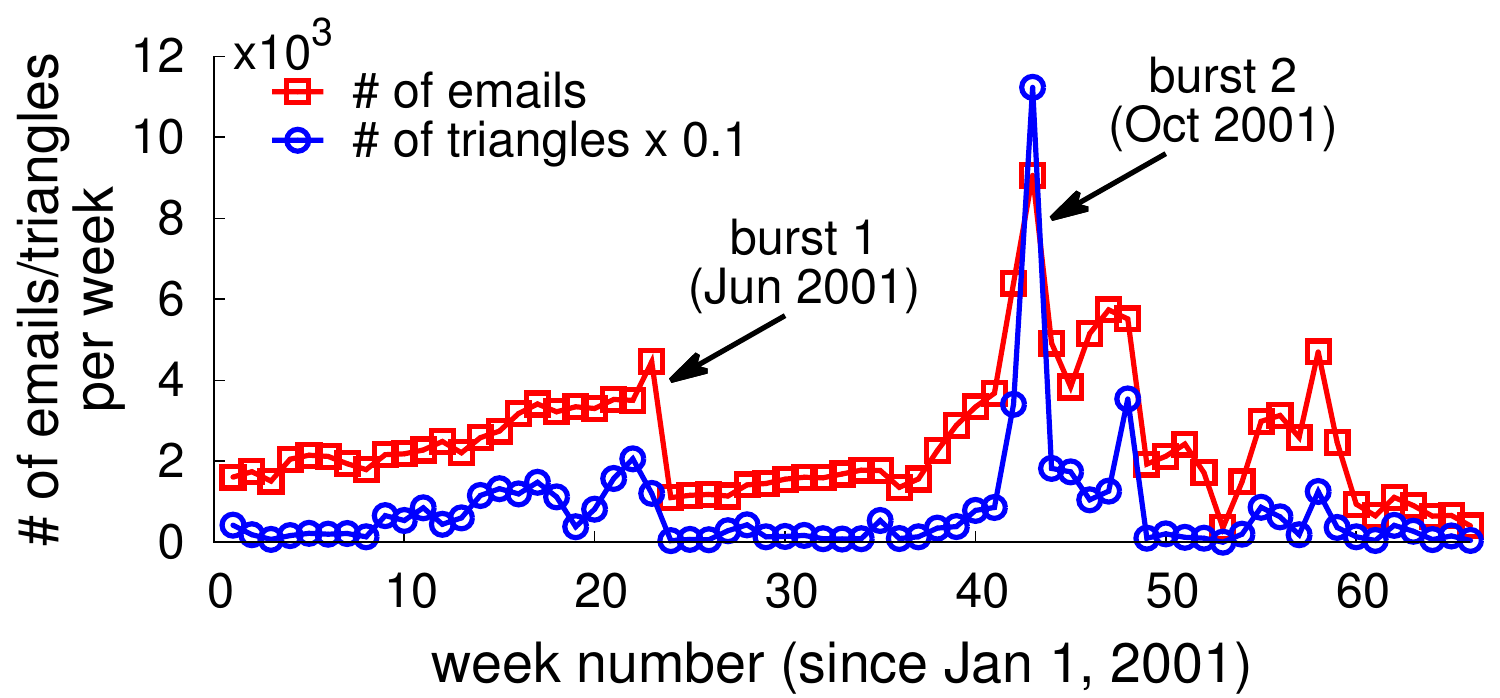}
\caption{Email and triangle volumes per week\label{fig:email_volume}}
\end{figure}

Because the data has been cleaned, the number of user-user interactions, i.e.,
number of sent emails per time window, reliably indicates burst occurrences.
We show the number of emails sent per week in Fig.~\ref{fig:email_volume}, and
observe at least two bursts that occurred in Jun and Oct 2001, respectively.
We also show the number of interaction triangles formed during each week.
The Pearson correlation coefficient (PCC) between the email and triangle volum
series is $0.8$, which reflects a very strong correlation.
The sudden increase (or decrease) of email volumes during the two bursts is
accompanied with the sudden increase (or decrease) of the number of triangles.
Thus, this observation verifies our claim that the emergence of a burst is
accompanied with the formation of triangles in networks.

\begin{figure}[htp]
\centering
\subfloat[Burst 1 ``shifts'' the dist. to left
  \label{fig:burst1}]{\includegraphics[width=.5\linewidth]{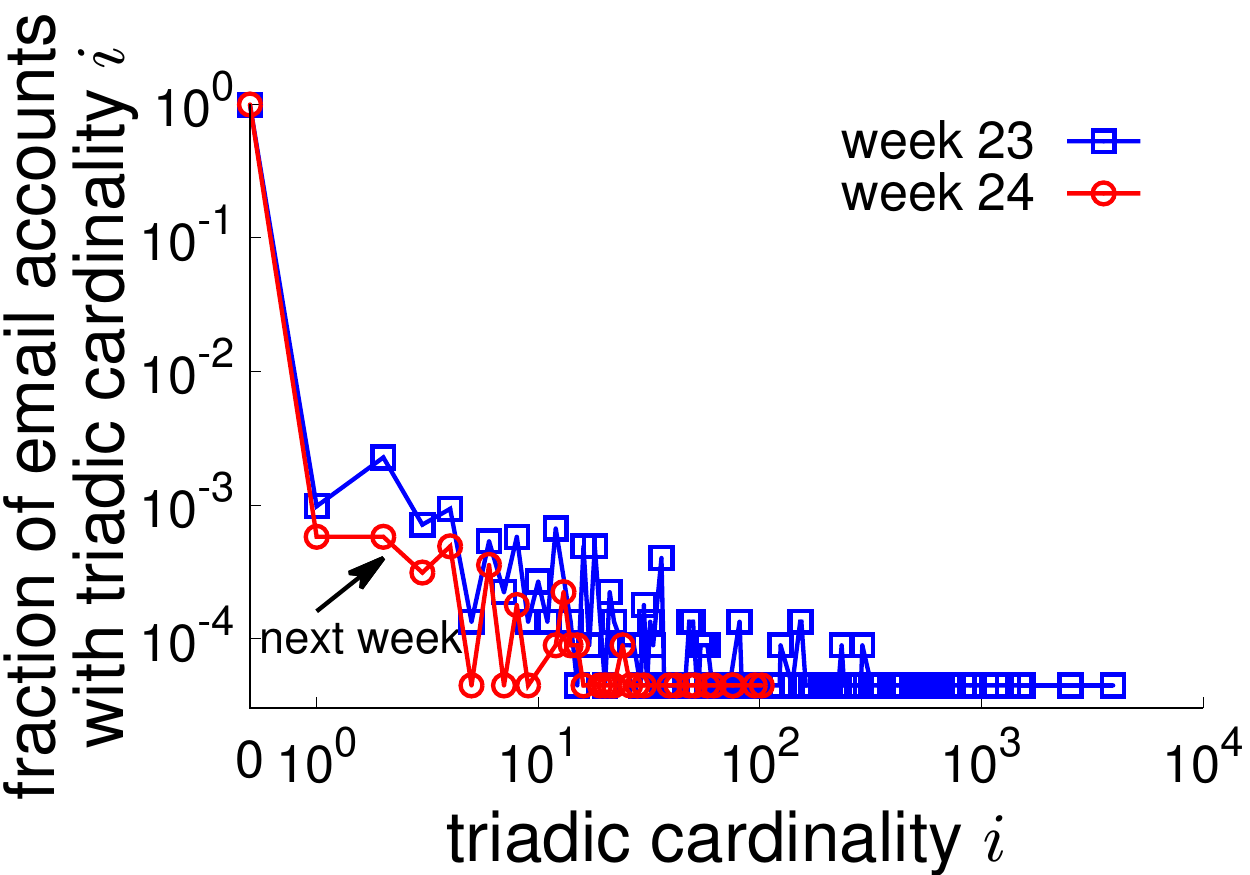}}
\subfloat[Burst 2 ``shifts'' the dist. to right
  \label{fig:burst2}]{\includegraphics[width=.5\linewidth]{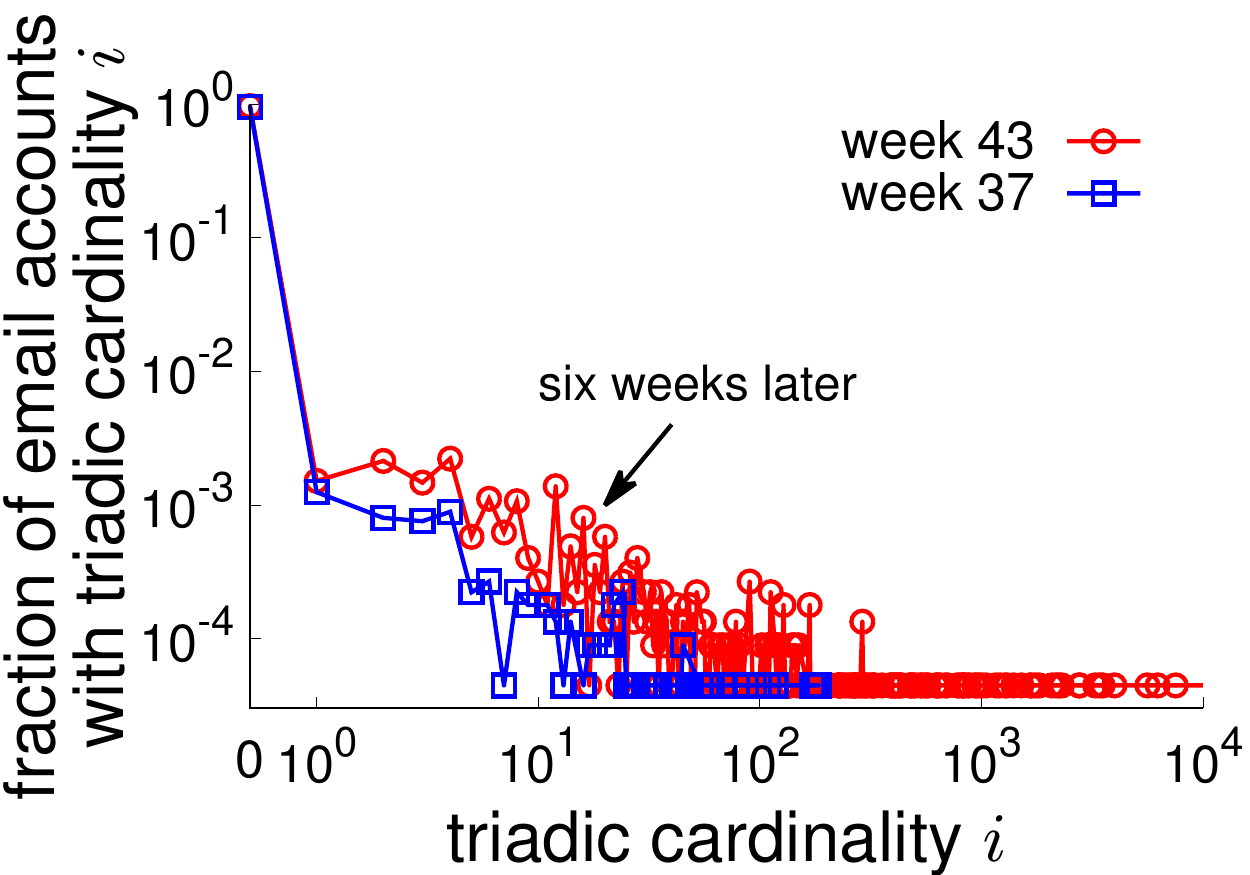}}
  \caption{Bursts change distribution curves.
    For burst 1 (burst 2), the probability mass at $i=0$ slightly increases
    (decreases) actually.\label{fig:burst}}
\end{figure}

\header{How bursts change triadic cardinality distributions.}
Our burst detection method relies on a claim that, when a burst occurs, the triadic
cardinality distribution changes.
To see this, we show the triadic cardinality distributions before and during the
bursts in Fig.~\ref{fig:burst}.
For the first burst, due to the sudden decrease of email communications from week
23 to week 24, we observe in Fig.~\ref{fig:burst1} that the distribution ``shifts''
to the left.
While for the second burst, due to the gradual increase of email communications, we
observe in Fig.~\ref{fig:burst2} that the distribution in week 43 shifts to the
right in comparison to previous weeks.
Again, the observation verifies our claim that triadic cardinality distribution
changes when a burst occurs.

\header{Impacts of spam.}
As we mentioned earlier, if spam exists, simply using the volume of user
interactions to detect bursts will result in false alarms, while the triadic
cardinality distribution is a good indicator immune to spam.
To demonstrate this claim, suppose a spammer suddenly becomes active in week 23,
and generates email spams to distort the original triadic cardinality distribution
of week 23.
We consider the following two spamming strategies:
\begin{itemize}
\item \emph{Random}: The spammer randomly chooses many target users to send spam.
\item \emph{Random-Friend}: At each step, the spammer randomly chooses a user and a
  random friend of the user\footnote{We assume two Enron users are friends if they
    have at least one email communication in the dataset.}, as two targets; and
  sends spams to each of these two targets.
  The spammer repeats this step a number of times.
\end{itemize}

\begin{figure}[htp]
\centering
\subfloat[KL divergence\label{fig:kl}]{%
  \includegraphics[width=.5\linewidth]{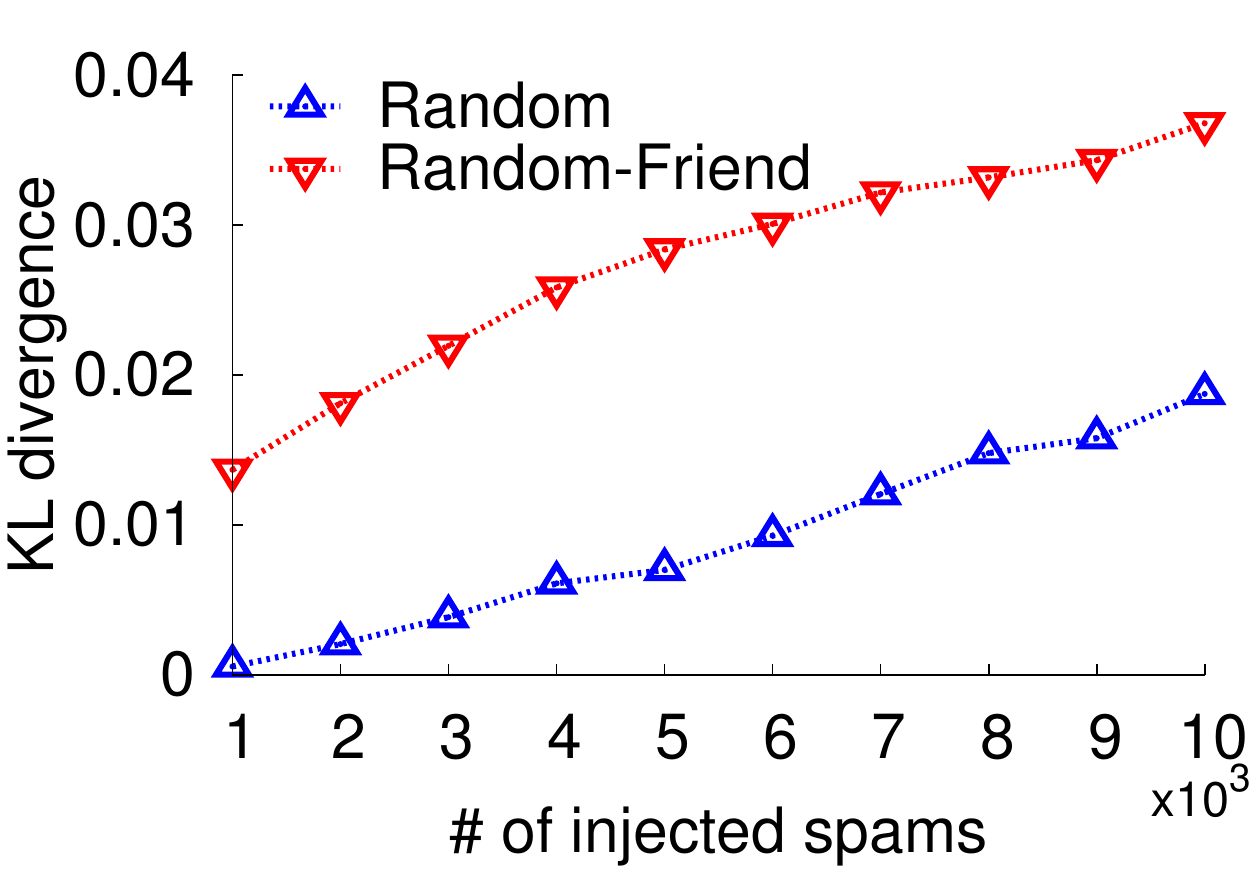}}
\subfloat[Distorted distributions\label{fig:distorted}]{%
  \includegraphics[width=.5\linewidth]{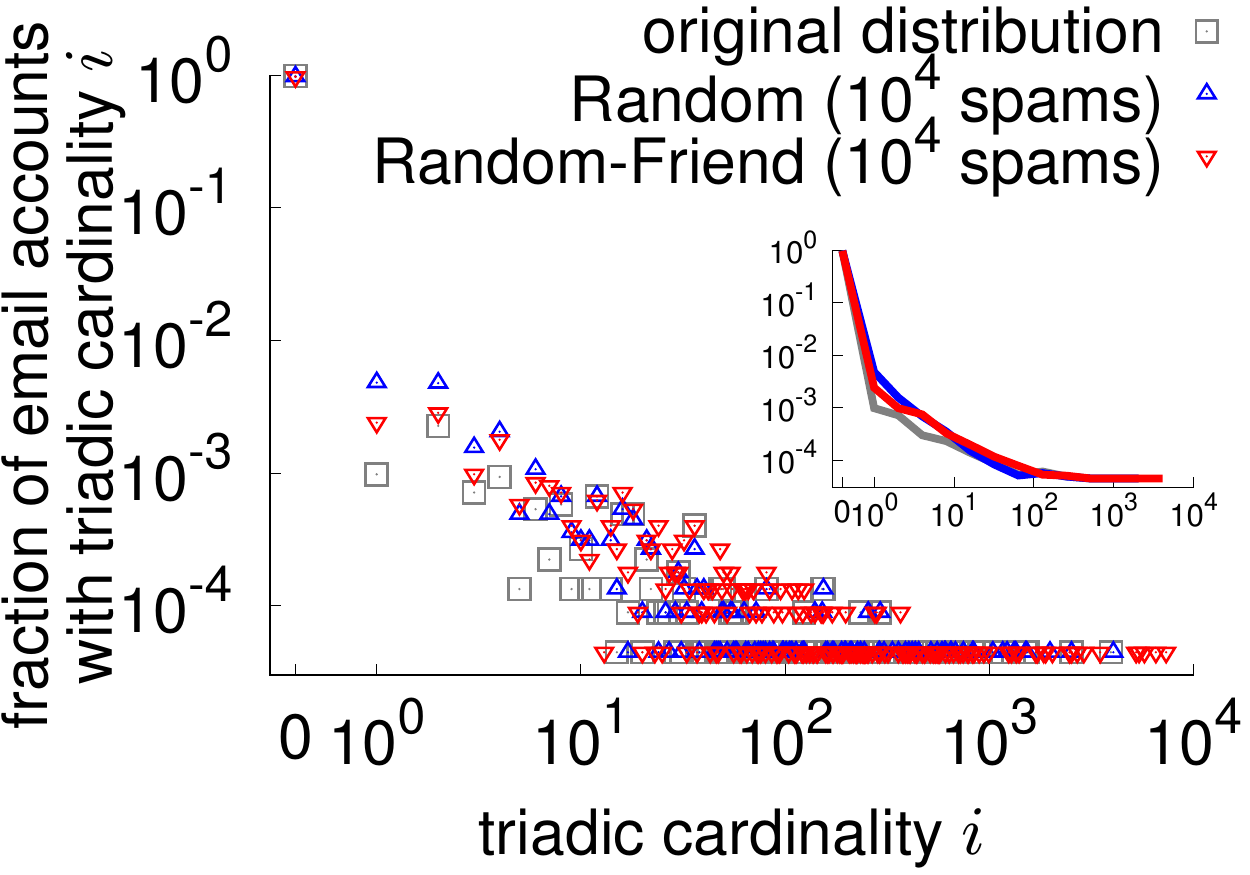}}
\caption{Impacts of spam.
  In (b), the inset shows fitted curves of the three distributions.}
\end{figure}

In order to measure the extent that spams can distort the original triadic
cardinality distribution of week 23, we use Kullback-Leibler (KL) divergence to
measure the difference between the original and distorted distributions.
The relationship between KL divergence and the number of injected spams is shown in
Fig.~\ref{fig:kl}.
For both strategies, KL divergences both increase as more spams are injected into
the interaction network, which is expected.
The Random-Friend strategy can cause larger divergences than the Random strategy,
as Random-Friend strategy is easier to introduce new triangles to the interaction
network of week 23 for the reason that two friends are more likely to communicate
in a week.
However, even when $10^4$ spams are injected, the spams incur an increasing KL
divergence of less than $0.04$.
From Fig.~\ref{fig:distorted}, we can see that the divergence is indeed small.
(This may be explained by the ``{\em center of attention}''
phenomenon~\cite{Backstrom2011}, i.e., a person may have hundreds of friends but he
usually only interacts with a small fraction of them in a time window.
Hence, Random-Friend strategy does not form many triangles.)
Therefore, these observations verify that triadic cardinality distribution is
robust against common spamming attacks.

\subsection{Validating Estimation Performance}

In the second experiment, we evaluate the MLE performance using different sampling
methods and demonstrate the computational efficiency.

\subsubsection{\textbf{Datasets}}
Because the input of our estimation methods is actually a sampled graph, we use
public available graphs of different types and scales from the SNAP graph
repository (\url{http://snap.stanford.edu/data}) as our testbeds.
We summarize the statistics of these graphs in Table~\ref{tab:graphs}.

\begin{table}[htp]
\centering
\caption{Network statistics\label{tab:graphs}}
\begin{tabular}{l|l|r|r}
\hline\hline
Network & Type                 & Nodes       & Edges        \\
\hline
HepTh   & directed, citation   & $27,770$    & $352,807$    \\
DBLP    & undirected, coauthor & $317,080$   & $1,049,866$  \\
YouTube & undirected, OSN      & $1,134,890$ & $2,987,624$  \\
Pokec   & directed, OSN        & $1,632,803$ & $30,622,564$ \\
\hline
\end{tabular}
\end{table}

For each graph, we first shuffle the edges to form a stream, then we apply stream
sampling methods on the stream, and obtain a sampled graph.
We calculate the triadic cardinality for nodes in the sampled graph, and obtain
statistics $g$.
Note that the estimator uses $g$ to obtain an estimate of the triadic cardinality
distribution for each graph, which is then compared with the ground truth
distribution, i.e., the triadic cardinality distribution of the original unsampled
graph, to evaluate the performance of the estimation method.

\subsubsection{\textbf{CRLB analysis}}

Our goal is to compare the amount of information contained in edge samples
collected using different sampling methods, in terms of CRLB.
A small CRLB indicates small asymptotic variance of a MLE, and hence implies that
the corresponding sampling method is efficient in gathering information from data.
We will mainly use the HepTh and DBLP networks in this study, and for ease of
conducting matrix algebra, we truncate the stream with $W=20$ by discarding edges
that may increase a node's triadic cardinality to larger than $20$.

\begin{figure}[t]
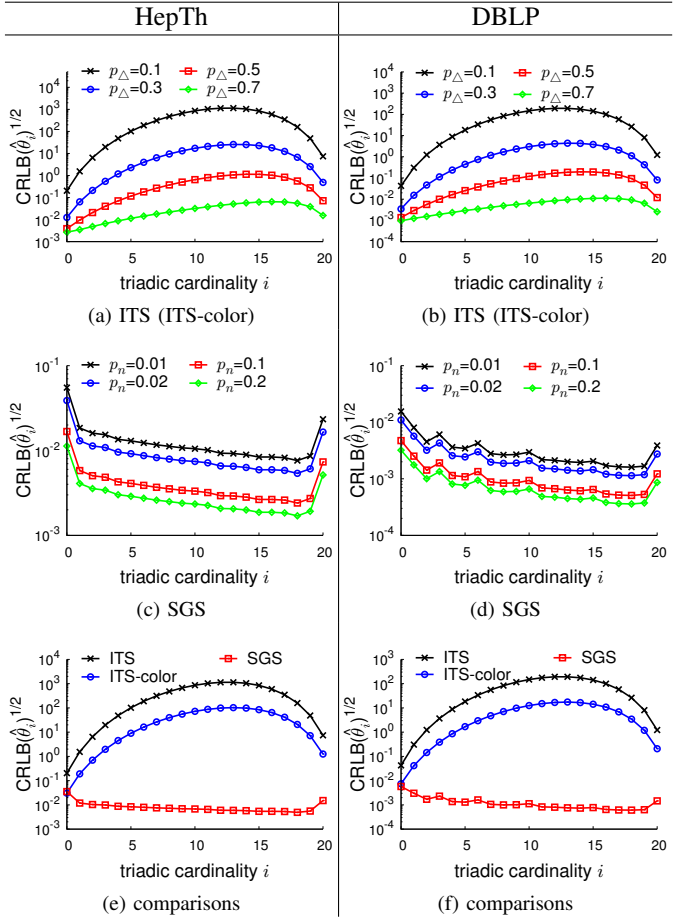

\centering
  \begin{tabular}{@{}c@{}|@{}c@{}}
    \hline
    HepTh & DBLP \\
    \hline
    \twofigs[f:crlb_its]{ITS (ITS-color)}{hepth_crlb_ITS}{dblp_crlb_ITS} \\
    \twofigs[f:crlb_sgs]{SGS}{hepth_crlb_SGS}{dblp_crlb_SGS} \\
    \twofigs[f:crlb_cmp]{comparisons}{hepth_crlb_cmp}{dblp_crlb_cmp} \\
  \end{tabular}
  \caption{CRLB analysis when graph size is known.
    $\alpha=0.1$.
    In~\protect\subref{f:crlb_cmp_1} and~\protect\subref{f:crlb_cmp_2},
    $\pt^\text{ITS}=0.1$ and $\pt^\text{ITS-color}=0.22$.
    In~\protect\subref{f:crlb_cmp_1}, $p_n=0.024$.
    In~\protect\subref{f:crlb_cmp_2}, $p_n=0.07$.}
  \label{fig:crlb}
\end{figure}

We depict the results when graph size is known in Fig.~\ref{fig:crlb}.
In~\subref{f:crlb_its_1} and~\subref{f:crlb_its_2}, we show the rooted CRLB of ITS
and ITS-color with different triangle sampling rates $\pt$ on the two networks
respectively.
As expected, when $\pt$ increases, CRLB decreases, indicating that we can obtain
more accurate estimates by increasing edge sampling rate.
However, we find that ITS and ITS-color are not efficient in gathering information
from data.
As we can see, to decrease CRLB to less than $0.1$, we need to increase $\pt$ to
about $0.6$, which corresponds to a very large edge sampling rate!
We then study the performance of SGS in~\subref{f:crlb_sgs_1}
and~\subref{f:crlb_sgs_2}.
We observe that CRLB decreases when $p_n$ increases, i.e., when more nodes (or
subgraphs) are sampled.
We also observe that CRLB of SGS is much smaller than ITS, even with small $p_n$.
It seems that SGS is more efficient in gathering information from data than ITS.
However, people may argue that SGS may sample more edges than ITS even with small
$p_n$.
To compare them fairly, we need to fix the number of edge samples used by different
methods.
In ITS or ITS-color, if the graph contains $m$ edges, then ITS or ITS-color samples
$mp$ edges on average.
In SGS, because a randomly chosen node has $\sum_i i\theta_i$ triangles, then
approximately, SGS samples $\Theta(np_n\sum_ii\theta_i)$ edges on average (if we
assume $\text{\#edges}=\Theta(\text{\#triangles})$).
In the experiment, we turn $\pt$ and $p_n$ to make sure that different methods
indeed use same amounts of edge samples approximately, and show the results
in~\subref{f:crlb_cmp_1} and~\subref{f:crlb_cmp_2}.
We can see clearly that SGS is indeed more efficient than ITS and ITS-color.
ITS-color is also more efficient than ITS since ITS-color samples a triangle with
larger probability than ITS using same edge sampling rate.

\begin{figure}[t]
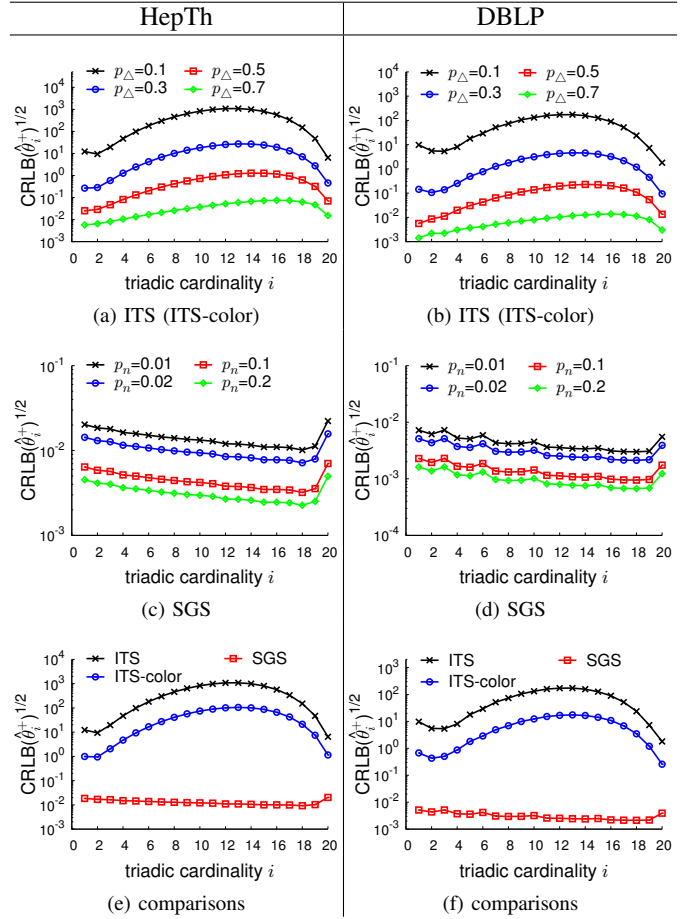

\centering
  \begin{tabular}{@{}c@{}|@{}c@{}}
    \hline
    HepTh & DBLP \\
    \hline
    \twofigs{ITS (ITS-color)}{hepth_crlb_un_ITS}{dblp_crlb_un_ITS} \\
    \twofigs{SGS}{hepth_crlb_un_SGS}{dblp_crlb_un_SGS} \\
    \twofigs[f:crlb_un_cmp]{comparisons}{hepth_crlb_un_cmp}{dblp_crlb_un_cmp} \\
  \end{tabular}
  \caption{CRLB analysis when graph size is unknown.
    In ITS and ITS-color, $\alpha=0.1$.
    Parameters in~\protect\subref{f:crlb_un_cmp_1}
    and~\protect\subref{f:crlb_un_cmp_2} are the same as in
    Fig.~\ref{fig:crlb}.\label{fig:crlb_un}}
\end{figure}

We also conduct same experiments when graph size is unknown, and the results are
depicted in Fig.~\ref{fig:crlb_un}.
The observations are consistent with the results when graph size is known.

\subsubsection{\textbf{NRMSE analysis}}

CRLB reflects the asymptotic variance of a MLE, i.e., the variance when sample size
approaches infinity.
However, in practice, we cannot collect infinite many samples because the number of
edges in a stream is finite, or afford to use large sample rate.
When sample rate is small, or collected edge samples are not large, MLE is usually
biased and we cannot leverage CRLB to analyze its performance
(see~\cite[p.~147]{Trees2001} for details).
Instead, we propose to use the normalized rooted mean squared error (NRMSE) of an
estimator, which is defined by
$\text{NRMSE}(\hat\theta_i)=\sqrt{\E{(\hat\theta_i-\theta_i)^2}}/\theta_i$.
The smaller the NRMSE, the more accurate an estimator is.
In the following experiment, we mainly use the HepTh network, and compare different
sampling methods using approximately the same amount of edge samples.

\begin{figure}[t]
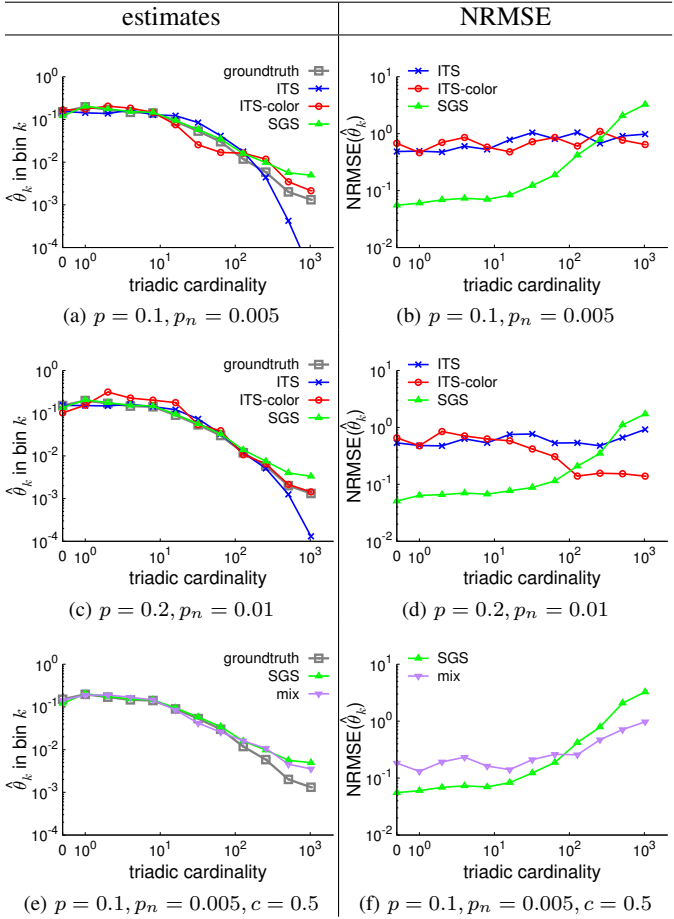

\centering
  \begin{tabular}{@{}c@{}|@{}c@{}}
    \hline
    estimates & NRMSE \\
    \hline
    \twofigs[f:nrmse_sm]{$p=0.1,p_n=0.005$}{db_est_p1}{db_nrmse_p1} \\
    \twofigs[f:nrmse_bg]{$p=0.2,p_n=0.01$}{db_est_p2}{db_nrmse_p2} \\
    \twofigs[f:nrmse_mx]{$p=0.1,p_n=0.005,c=0.5$}{db_est_p1_mix}
    {db_nrmse_p1_mix}\\
  \end{tabular}
  \caption{Estimation accuracy analysis on HepTh when graph size is known.}
  \label{fig:nrmse}
\end{figure}

We depict the results when graph size is known in Fig.~\ref{fig:nrmse}.
In~\subref{f:nrmse_sm_1} and~\subref{f:nrmse_sm_2}, we compare the estimates and
NRMSE of different methods.
In general, SGS is better than ITS-color, and ITS-color is better than ITS.
This observation is consistent with our previous CRLB analysis.
The NRMSE plots in~\subref{f:nrmse_sm_2} provide more valuable observations.
We observe that SGS can provide more accurate estimates for nodes with small
triadic cardinalities than ITS and ITS-color; however, SGS performs much worse than
ITS and ITS-color for nodes with large triadic cardinalities.
This observation can be explained from the different nature between SGS and ITS
based methods.
The ITS based methods sample each triangle with identical probability, and if
dependence between triangles is negligible, the sampling will be strongly biased
towards nodes with many triangles.
That is, nodes with larger triadic cardinalities are more likely to be sampled, and
for nodes with small triadic cardinalities, the triangles these nodes belonging to
will be seldom sampled.
This results in that triangles of small triadic cardinality nodes are difficult to
be sampled, and hence incurs large estimation error for these nodes.
SGS is completely different, and it reserves all the triangles of each node sample.
Because nodes are sampled with same probability, node samples will be dominated by
nodes with small triadic cardinalities.
Hence, SGS can provide more accurate estimates for the head of triadic cardinality
distribution.
However, SGS is inefficient in sampling nodes with large triadic cardinalities,
resulting in large NRMSE at the tail of the triadic cardinality distribution.
To address their weaknesses, one way is to increase sampling rates, as depicted
in~\subref{f:nrmse_bg_1} and~\subref{f:nrmse_bg_2}.
We observe that after increasing sampling rates, estimation accuracy increases more
or less for each method.
An alternative way is to design a mixture estimator, which combines the advantage
(and the disadvantage) of each method.
For example, we define
\[
  \hat\theta_i^\text{mix}\triangleq
  c\hat\theta_i^\text{ITS-color} + (1-c)\hat\theta_i^{SGS},
\]
where $c\in[0,1]$ is a constant.
$\hat\theta_i^\text{mix}$ has the property that, its variance is smaller than
$\hat\theta^\text{SGS}$ for nodes with large triadic cardinalities, with the loss
of accuracy for nodes with small triadic cardinalities, and the variance of the
mixture estimator achieves minimal at $c^* = var(\hat\theta_i^{SGS}) /
[var(\hat\theta_i^\text{ITS-color}) + var(\hat\theta_i^{SGS})]$.
Figs.~\subref{f:nrmse_mx_1} and~\subref{f:nrmse_mx_2} show the results of a mixture
estimator with $c=0.5$.
We indeed observe improvements for nodes with large triadic cardinalities.

\begin{figure}[t]
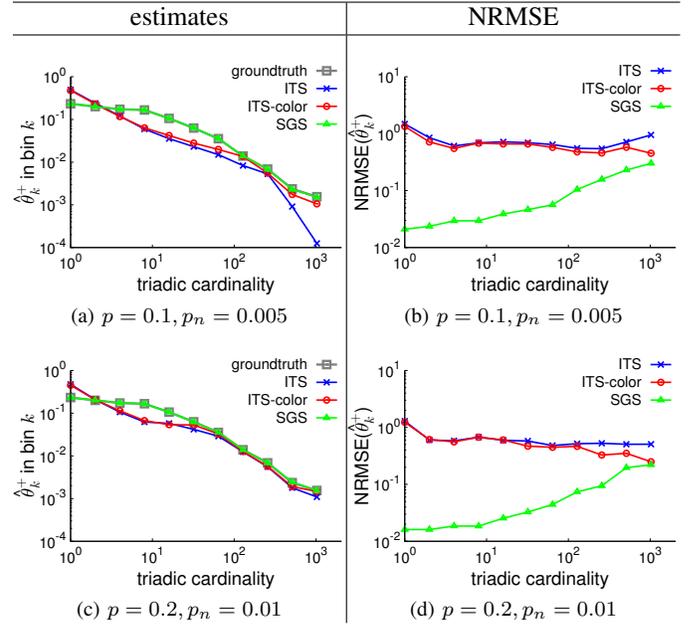

\centering
  \begin{tabular}{@{}c@{}|@{}c@{}}
    \hline
    estimates & NRMSE \\
    \hline
    \twofigs{$p=0.1,p_n=0.005$}{db_est_un_p1}{db_nrmse_un_p1} \\
    \twofigs{$p=0.2,p_n=0.01$}{db_est_un_p2}{db_nrmse_un_p2} \\
  \end{tabular}
  \caption{Estimation accuracy analysis on HepTh when graph size is unknown.}
  \label{fig:nrmse_un}
\end{figure}

We also conduct experiments when graph size is unknown,
and show the results in Fig.~\ref{fig:nrmse_un}.
The observations are consistent with the results when graph size is known in
general, and we observe that SGS has smaller NRMSE than ITS based methods even for
nodes with large triadic cardinalities.

\begin{figure}[htp]
  \centering
  \subfloat[YouTube]{\includegraphics[width=.5\linewidth]{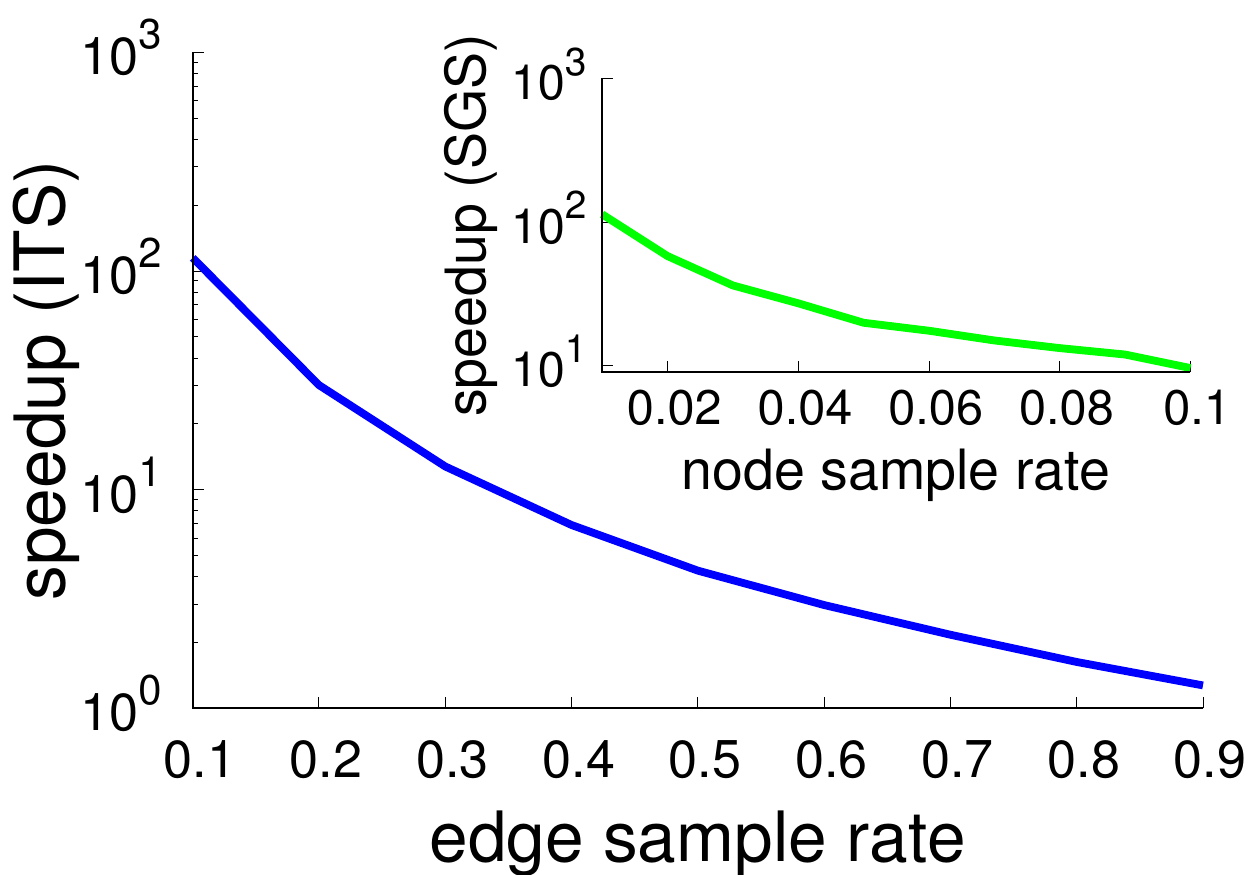}}
  \subfloat[Pokec]{\includegraphics[width=.5\linewidth]{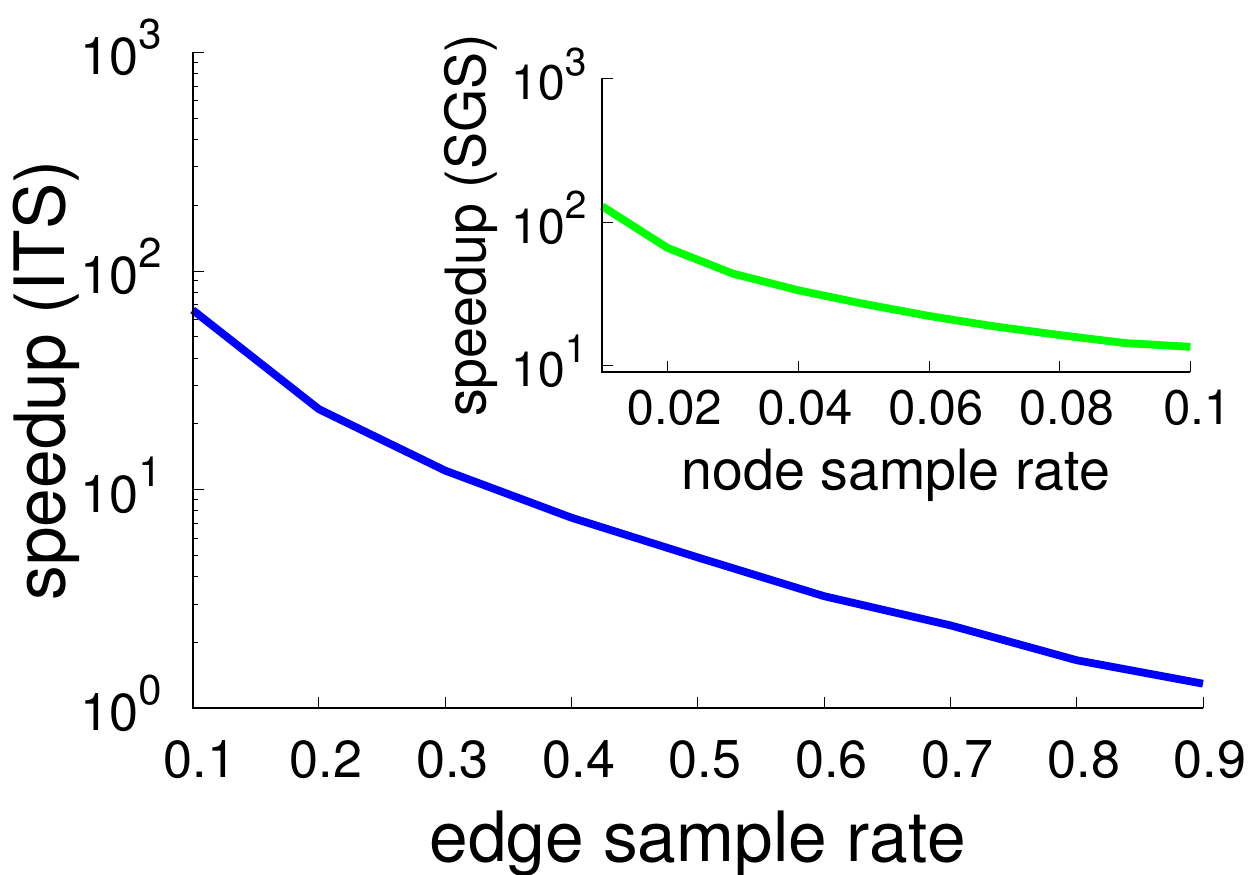}}
  \caption{Computational efficiency comparison}
  \label{fig:spd}
\end{figure}

Finally, we also conducted experiments on two larger networks, YouTube and Pokec.
Here, we mainly compare the computational efficiency of our sampling approach
against a naive method that uses all of the original graph to calculate $\theta$ in
an exact fashion.
The results are depicted in Fig.~\ref{fig:spd}.
In general, for ITS, using edge sampling rates between $0.1$ and $0.3$, we have
speedup about $10$ to $50$.

\subsection{Application: Tracking Triadic Cardinality Distributions
  during the 2014 Hong Kong Occupy Central Movement}

Last, we conduct an application to illustratively show the usefulness of tracking
triadic cardinality distributions during the 2014 Hong Kong Occupy Central movement
in Twitter.

\header{Hong Kong Occupy Central movement} a.k.a.~the Umbrella Revolution, began in
Sept 2014 when activists in Hong Kong protested against the government and occupied
several major streets of Hong Kong to go against a decision made by China's
Standing Committee of the National People's Congress on the proposed electoral
reform.
Protesters began gathering from Sept 28 on and we collected the data between Sept 1
and Nov 30 in 2014.

\header{Building a Twitter social activity stream.}
The input of our solution is a social activity stream from Twitter.
For Twitter itself, this stream is easily obtained by directly aggregating tweets
of users.
While for third parties who do not own user's tweets, the stream can be obtained by
following a set of users, and aggregating tweets from these users to form a social
activity stream.
Since the movement had already begun prior to our starting this work, we rebuilt
the social activity stream by searching tweets containing at least one of the
following hashtags: \#OccupyCentral, \#OccupyHK, \#UmbrellaRevolution,
\#UmbrellaMovement and \#UMHK, between Sept 1 and Nov 30 using Twitter search APIs.
This produced $66,589$ Twitter users, and these users form the detectors from whom
we want to detect bursts.
Next, we collect each user's tweets between Sept 1 and Nov 30, and extract user
mentions (i.e., user-user interactions) and user hashtags (i.e., user-content
interactions) from tweets to form a social activity stream, with a time span of
$91$ days.

\header{Settings.}
We set the length of a time window to be one day.
For interaction bursts caused by user-user interactions, because we know the user
population, i.e., $n=66,589$, we apply the first estimation method to obtain
$\hat\theta=(\hat\theta_0,\ldots,\hat\theta_W)$ for each window.
For cascading bursts caused by user-content interactions, as we do not know the
number of hashtags in advance, we apply the second method to obtain estimates
$\hat{n}_+$, i.e., the number of hashtags with at least one influence triangle, and
$\hat\vartheta^+=(\hat\vartheta_1^+,\ldots,\hat\vartheta_W^+)$ for each window.
Combining $\hat{n}_+$ with $\hat\vartheta^+$, we use $\hat{n}_+\hat\vartheta^+$,
i.e., frequencies, to characterize patterns of user-content interactions in each
window.

\begin{figure}[htp]
  \centering
  \subfloat[Interaction burst]{\includegraphics[width=.5\linewidth]{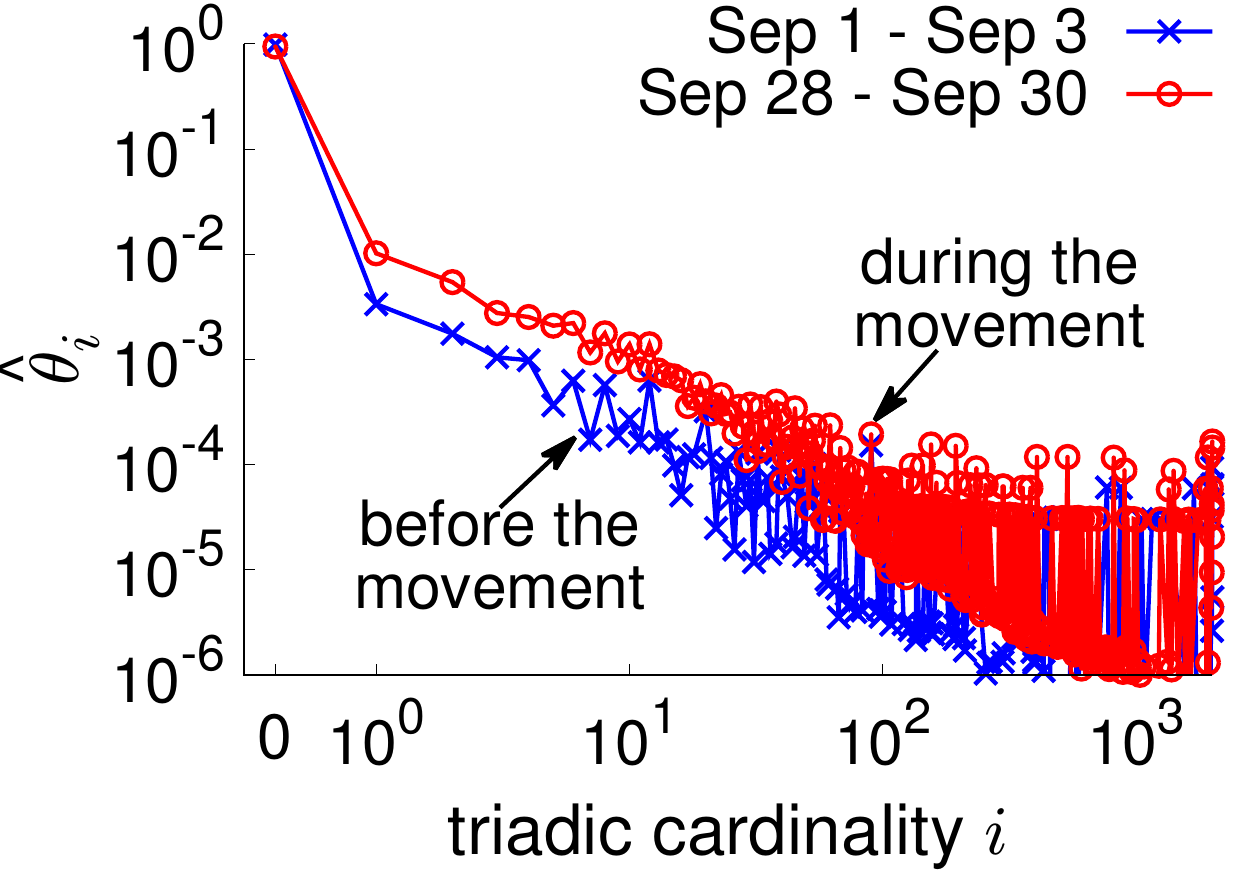}}
  \subfloat[Cascading burst]{\includegraphics[width=.5\linewidth]{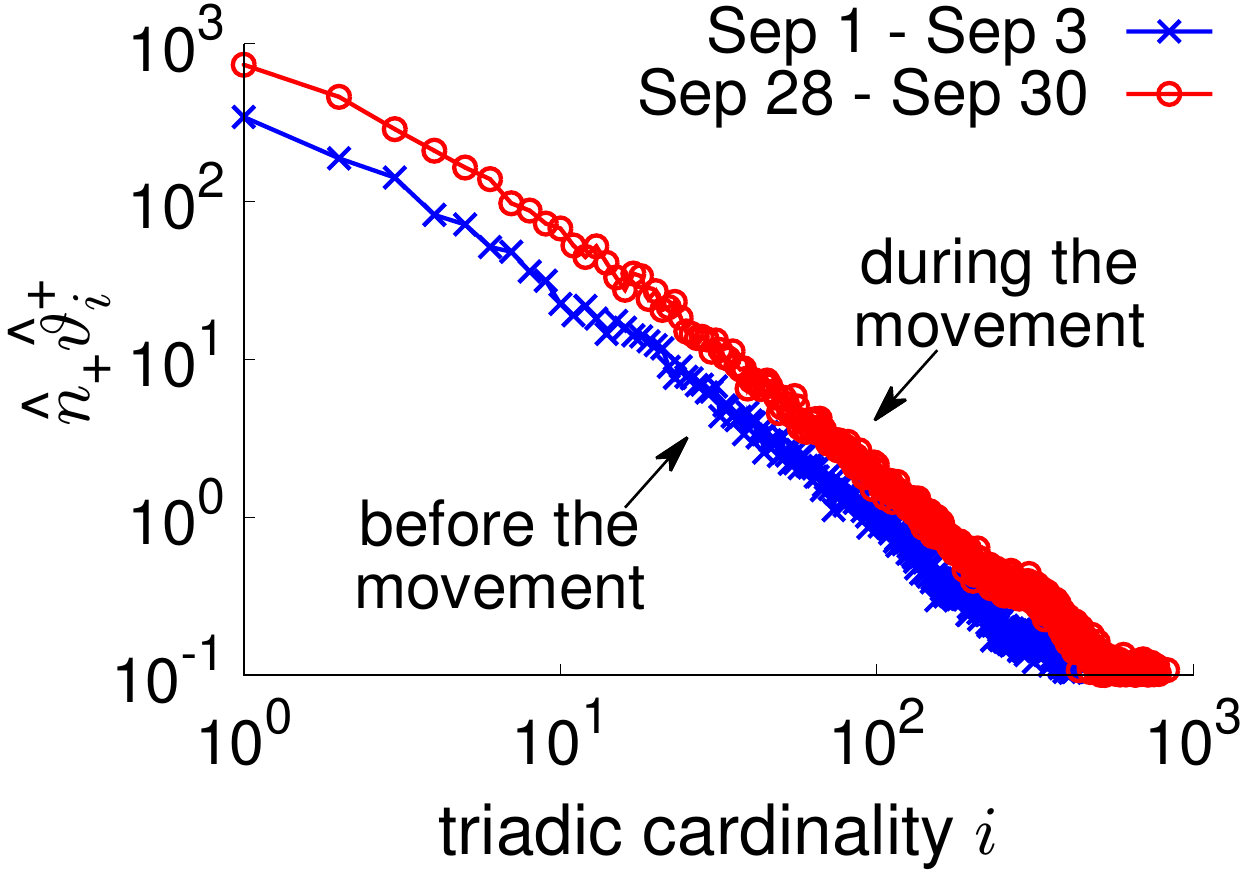}}
  \caption{Triadic cardinality distributions before and during the movement.
    Estimated using a mixture estimator with $c=0.5, p=0.2, p_n=0.002$.}
  \label{fig:diff_week}
\end{figure}

\header{Results.}
We first answer the question: are there significant differences for the two
distributions before and during the movement?
In Fig.~\ref{fig:diff_week}, we compare the distributions before (Sept 1 to Sept 3)
and during (Sept 28 to Sept 30) the movement.
We can find that when the movement began on Sept 28, the distributions of the two
kinds of interactions shift to the right, indicating that many interaction and
influence triangles form when the movement starts.
Therefore, these observations confirm our motivation for detecting bursts by
tracking triadic cardinality distributions.

\begin{figure*}[htp]
  \centering
  \begin{tikzpicture}
    \node at (0,0) {\includegraphics[width=.95\linewidth]{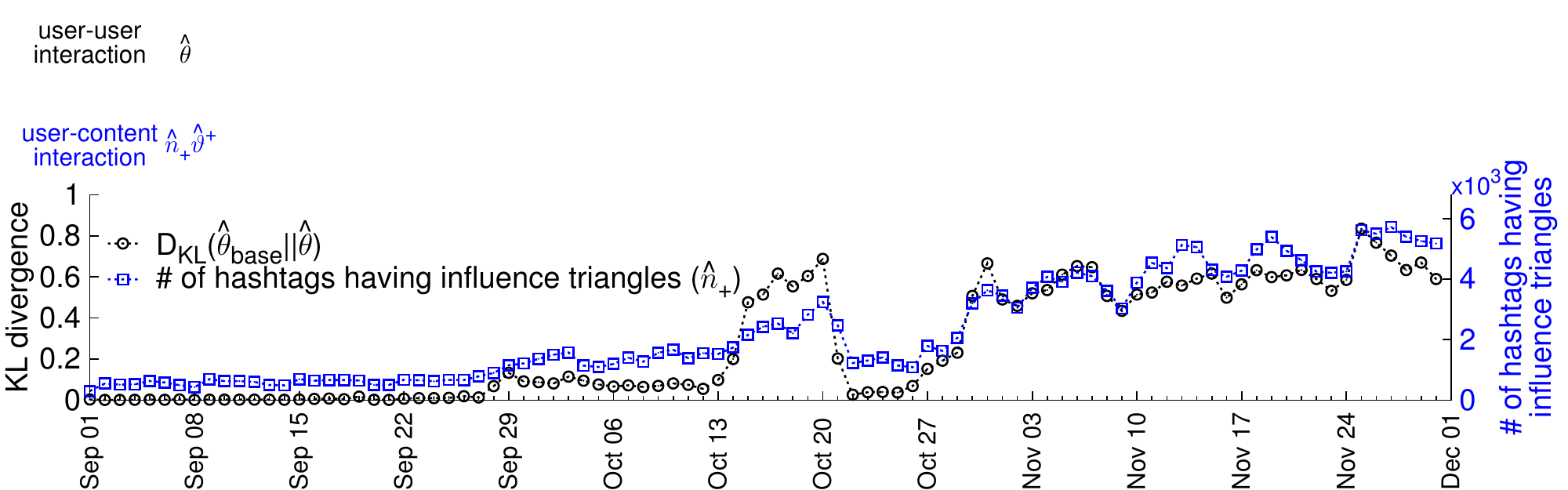}};
    \node at (0.4,1.45) {\includegraphics[width=.65\linewidth]{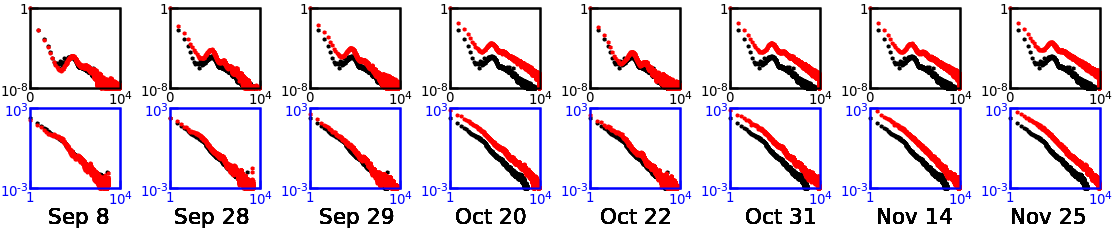}};
  \end{tikzpicture}
  \caption{Triadic cardinality distributions change during the 2014 Hong Kong
    Occupy Central movement in Twitter.
  }
  \label{fig:kl_hk}
\end{figure*}

Next, we track the daily triadic cardinality distributions to look up the
distribution change during the movement.
To characterize the sudden change in the distributions, we use KL divergence to
calculate the difference between $\hat\theta$ and a base distribution
$\theta_\text{base}$.
The base distribution $\theta_\text{base}$ represents a distribution when the
network is dormant, i.e., no bursts are occurring.
Here we omit the technique details, and simply average the triadic cardinality
distributions from Sept 1 to Sept 7 to obtain an approximate base distribution
$\hat\theta_\text{base}$, and show the KL divergence
$D_\text{KL}(\hat\theta_\text{base}\parallel\hat\theta)$ in Fig.~\ref{fig:kl_hk}.

We find that the KL divergence exhibits a sudden increase on Sept 28 when the
movement broke out.
The movement keeps going on and reaches a peak on Oct 19 when repeated clashes
happened in Mong Kok at that time.
The movement temporally returned to peace between Oct 22 and Oct 25, and restarted
again after Oct 26.
In Fig.~\ref{fig:kl_hk}, we also show the estimated number of hashtags having at
least one influence triangle.
Its trend is similar to the trend of KL divergence which indicates that the
movement is accompanied with rumors spreading in a word-of-mouth manner.

In conclusion, the application in this section demonstrates that the using of the
triadic cardinality distribution can track bursts from a social activity stream and
the result is consistent with real world events.

\section{Related Work}
\label{sec:relatedwork}

Kleinberg first studied the topic of burst detection from streams
in~\cite{Kleinberg2002}, where he used a multistate automaton to model a stream
consisting of messages, e.g., an email stream.
The occurrence of a burst is modeled by an underlying state transiting into a
bursty state that emits messages at a higher rate than at the non-bursty state.
Based on this model, many variant models are proposed for detecting bursts from
document streams~\cite{Yi2005,Mathioudakis2010}, e-commerce
queries~\cite{Parikh2008}, time series~\cite{Zhu2003}, and social
networks~\cite{Eftekhar2013}.
Although these models are theoretically interesting, some assumptions made by them
are inappropriate, such as the Poisson process of message arrivals
(see~\cite{Barabasi2005}) and nonexistence of spams/bots, which may limit their
practical usage.

The topic of (anomaly) event detection is also related to our work.
Recently, Chierichetti et al.~\cite{Chierichetti2014} found that Twitter user
tweeting and retweeting count information can be used to detect sub-events during
some large event such as the soccer World Cup of 2010.
Takahashi et al.~\cite{Takahashi2011} proposed a probabilistic model to detect
emerging topics in Twitter by assigning an anomaly score for each user.
Sakaki et al.~\cite{Sakaki2010} proposed a spatiotemporal model to detect
earthquakes using tweets.
Manzoor et al.~\cite{Manzoor2016} studied anomaly event detection from a graph
stream based on graph similarity metrics.
Different from theirs, we exploit the triangle structure existing in user
interactions which is robust against common spams and can be efficiently estimated
using our method.

The triangle structure can be considered as a type of network motif, which is
introduced in~\cite{Milo2002} when the authors were studying how to characterize
structures of different types of networks.
Turkett et al.~\cite{Turkett2011} used motifs to analyze computer network usage,
and~\cite{Wang2014} proposed sampling methods to efficiently estimate motif
statistics in a large graph.
However, both the motivation in~\cite{Turkett2011} and subgraph statistics defined
in~\cite{Wang2014} are different from ours.

Recently, there are many works on estimating the number of global and local
triangles~\cite{Tsourakakis2009,Budak2011,Pavan2013,Jha2013,Ahmed2014,Lim2015a,Stefani2016a,Wu2016b},
or clustering coefficient~\cite{Seshadhri2013} in a large graph.
However, triadic cardinality distribution is much complicated than triangle counts,
and these methods cannot be used to estimate the triadic cardinality distribution.
Becchetti et al.~\cite{Becchetti2008} used a min-wise hashing method to
approximately count triangles for each individual node in an undirected simple
graph.
Our method does not rely on counting triangles for each individual node.
Rather, we use a carefully designed estimator to estimate the statistics from a
sampled graph, which is demonstrated to be efficient and accurate.

\section{Conclusion}
\label{sec:conclusion}

Online social networks provide various ways for users to interact with other users
or media content over the Internet, which bridge the online and offline worlds
tightly.
This provides an opportunity to researchers to leverage online users' interactions
to detect bursts that may cause negative impacts to the offline world.
This work studied the burst detection problem from a high-speed social activity
stream generated by user's interactions in an OSN.
We show that the emergence of bursts caused by either user-user or user-content
interaction are accompanied with the formation of triangles in users' interaction
networks.
This finding prompts us to devise a novel method for burst detection in OSNs by
introducing the triadic cardinality distribution.
Triadic cardinality distribution is found to be robust against common spamming
attacks which makes it a more suitable indicator for detecting bursts than the
volume of user activities.
We design a sample-estimate solution that aims to estimate triadic cardinality
distribution from a sampled social activity stream.
We show that, in general, SGS is more efficient in gathering information from data
than ITS based methods.
However, SGS incurs larger NRMSE than ITS for nodes with large triadic
cardinalities.
We can combine ITS and SGS and use a mixture estimator to further reduce the NRMSE
of SGS at the tail estimates.
We believe our work sets the foundation for robust burst detection, and it is an
open problem for finding and designing better or optimal sampling and estimation
methods.

\ifCLASSOPTIONcaptionsoff
  \newpage
\fi

\bibliographystyle{IEEEtran}

\begin{thebibliography}{10}
\providecommand{\url}[1]{#1}
\csname url@samestyle\endcsname
\providecommand{\newblock}{\relax}
\providecommand{\bibinfo}[2]{#2}
\providecommand{\BIBentrySTDinterwordspacing}{\spaceskip=0pt\relax}
\providecommand{\BIBentryALTinterwordstretchfactor}{4}
\providecommand{\BIBentryALTinterwordspacing}{\spaceskip=\fontdimen2\font plus
\BIBentryALTinterwordstretchfactor\fontdimen3\font minus
  \fontdimen4\font\relax}
\providecommand{\BIBforeignlanguage}[2]{{%
\expandafter\ifx\csname l@#1\endcsname\relax
\typeout{** WARNING: IEEEtran.bst: No hyphenation pattern has been}%
\typeout{** loaded for the language `#1'. Using the pattern for}%
\typeout{** the default language instead.}%
\else
\language=\csname l@#1\endcsname
\fi
#2}}
\providecommand{\BIBdecl}{\relax}
\BIBdecl

\bibitem{Harvey2009}
M.~Harvey, ``Fans mourn artist for whom it didn{\textquoteright}t matter if you
  were black or white,''
  \url{http://wayback.archive.org/web/20100531165925/http://www.timesonline.co.uk/tol/news/world/us_and_americas/article6580897.ece},
  Retrived Jun 2017.

\bibitem{Shiels2009}
M.~Shiels, ``Web slows after {Jackson}'s death,''
  \url{http://news.bbc.co.uk/2/hi/technology/8120324.stm}, Retrived Jun 2017.

\bibitem{LondonRiots}
``London riots: More than 2,000 people arrested over disorder,''
  \url{http://www.mirror.co.uk/news/uk-news/london-riots-more-than-2000-people-185548},
  Retrived Jun 2017.

\bibitem{Chu2010}
Z.~Chu, S.~Gianvecchio, H.~Wang, and S.~Jajodia, ``Who is tweeting on
  {Twitter}: Human, bot, or cyborg?'' in \emph{Proceedings of the 26th Annual
  Computer Security Applications Conference}, 2010.

\bibitem{Grier2010}
C.~Grier, K.~Thomas, V.~Paxson, and M.~Zhang, ``@spam: The underground on 140
  characters or less,'' in \emph{Proceedings of the ACM SIGSAC Conference on
  Computer and Communications Security}, 2010.

\bibitem{Stringhini2010a}
G.~Stringhini, C.~Kruegel, and G.~Vigna, ``Detecting spammers on social
  networks,'' in \emph{Proceedings of the 26th Annual Computer Security
  Applications Conference}, 2010.

\bibitem{Boshmaf2011}
Y.~Boshmaf, I.~Muslukhov, K.~Beznosov, and M.~Ripeanu, ``The socialbot network:
  When bots socialize for fame and money,'' in \emph{Proceedings of the 27th
  Annual Computer Security Applications Conference}, 2011.

\bibitem{Thomas2011}
K.~Thomas, C.~Grier, V.~Paxson, and D.~Song, ``Suspended accounts in
  retrospect: An analysis of {Twitter} spam,'' in \emph{Proceedings of the 11th
  ACM SIGCOMM Conference on Internet Measurement}, 2011.

\bibitem{Beutel2013a}
A.~Beutel, W.~Xu, V.~Guruswami, C.~Palow, and C.~Faloutsos, ``{CopyCatch}:
  Stopping group attacks by spotting lockstep behavior in social networks,'' in
  \emph{Proceedings of the 22nd International World Wide Web Conference}, 2013.

\bibitem{Kleinberg2002}
J.~Kleinberg, ``Bursty and hierarchical structure in streams,'' in
  \emph{Proceedings of the 8th ACM SIGKDD International Conference on Knowledge
  Discovery and Data Mining}, 2002.

\bibitem{Yi2005}
J.~Yi, ``Detecting buzz from time-sequenced document streams,'' in
  \emph{Proceedings of the IEEE International Conference on e-Technology,
  e-Commerce and e-Service}, 2005.

\bibitem{Parikh2008}
N.~Parikh and N.~Sundaresan, ``Scalable and near real-time burst detection from
  ecommerce queries,'' in \emph{Proceedings of the 14th ACM SIGKDD
  International Conference on Knowledge Discovery and Data Mining}, 2008.

\bibitem{Eftekhar2013}
M.~Eftekhar, N.~Koudas, and Y.~Ganjali, ``Bursty subgraphs in social
  networks,'' in \emph{Proceedings of the 6th International ACM Conference on
  Web Search and Data Mining}, 2013.

\bibitem{Watts1998}
D.~J. Watts and S.~H. Strogatz, ``Collective dynamics of `small-world'
  networks,'' \emph{Nature}, vol. 393, pp. 440--442, 1998.

\bibitem{Kossinets2006}
G.~Kossinets and D.~J. Watts, ``Empirical analysis of an evolving social
  network,'' \emph{Science}, vol. 311, no. 5757, pp. 88--90, 2006.

\bibitem{Leskovec2007b}
J.~Leskovec, M.~McGlohon, C.~Faloutsos, N.~Glance, and M.~Hurst, ``Cascading
  behavior in large blog graphs,'' in \emph{Proceedings of the 7th SIAM
  International Conference on Data Mining}, 2007.

\bibitem{Rodrigues2011}
T.~Rodrigues, F.~Benevenuto, M.~Cha, K.~P. Gummadi, and V.~Almeida, ``On
  word-of-mouth based discovery of the web,'' in \emph{Proceedings of the 11th
  ACM SIGCOMM Conference on Internet Measurement}, 2011.

\bibitem{Tsotsis2011}
A.~Tsotsis, ``First credible reports of {Bin Laden}'s death spread like
  wildfire on {Twitter},''
  \url{https://techcrunch.com/2011/05/01/news-of-osama-bin-ladens-death-spreads-like-wildfire-on-twitter},
  Retrived Jun 2017.

\bibitem{Krikorian2013}
R.~Krikorian, ``New tweets per second record, and how!''
  \url{https://blog.twitter.com/engineering/en_us/a/2013/new-tweets-per-second-record-and-how.html},
  Retrived Jun 2017.

\bibitem{Pagh2012}
R.~Pagh and C.~E. Tsourakakis, ``Colorful triangle counting and a {MapReduce}
  implementation,'' \emph{Journal of Information Processing Letters}, vol. 112,
  no.~7, pp. 277--281, 2012.

\bibitem{Ahmed2014}
N.~K. Ahmed, N.~Duffield, J.~Neville, and R.~Kompella, ``Graph sample and hold:
  A framework for big-graph analytics,'' in \emph{Proceedings of the 20th ACM
  SIGKDD International Conference on Knowledge Discovery and Data Mining},
  2014.

\bibitem{Tsourakakis2009}
C.~E. Tsourakakis, U.~Kang, G.~L. Miller, and C.~Faloutsos, ``{DOULION}:
  Counting triangles in massive graphs with a coin,'' in \emph{Proceedings of
  the 15th ACM SIGKDD International Conference on Knowledge Discovery and Data
  Mining}, 2009.

\bibitem{Gjoka2011}
M.~Gjoka, M.~Kurant, C.~T. Butts, and A.~Markopoulou, ``Practical
  recommendations on crawling online social networks,'' \emph{IEEE Journal on
  Selected Areas in Communications}, vol.~29, no.~9, pp. 1872--1892, 2011.

\bibitem{Duffield2003}
N.~Duffield, C.~Lund, and M.~Thorup, ``Estimating flow distributions from
  sampled flow statistics,'' in \emph{Proceedings of the ACM Special Interest
  Group on Data Communication}, 2003.

\bibitem{Ribeiro2006}
B.~Ribeiro, D.~Towsley, T.~Ye, and J.~C. Bolot, ``Fisher information of sampled
  packets: An application to flow size estimation,'' in \emph{Proceedings of
  the 6th ACM SIGCOMM Conference on Internet Measurement}, 2006.

\bibitem{Tune2011}
P.~Tune and D.~Veitch, ``Fisher information in flow size distribution
  estimation,'' \emph{IEEE Transactions on Information Theory}, vol.~57,
  no.~10, pp. 7011--7035, 2011.

\bibitem{Wang2014a}
P.~Wang, X.~Guan, J.~Zhao, J.~Tao, and T.~Qin, ``A new sketch method for
  measuring host connection degree distribution,'' \emph{IEEE Transactions on
  Information Forensics and Security}, vol.~9, no.~6, pp. 948--960, 2014.

\bibitem{Veitch2015}
D.~Veitch and P.~Tune, ``Optimal skampling for the flow size distribution,''
  \emph{IEEE Transactions on Information Theory}, vol.~61, no.~6, pp.
  3075--3099, 2015.

\bibitem{Beta-bin}
``Beta-binomial distribution,''
  \url{https://en.wikipedia.org/wiki/Beta-binomial_distribution}, Retrived Jun
  2017.

\bibitem{Yu2002}
C.~Yu and D.~Zelterman, ``Sums of dependent {Bernoulli} random variables and
  disease clustering,'' \emph{Statistics and Probability Letters}, vol.~57,
  no.~1, pp. 363--373, 2002.

\bibitem{Durand2003}
M.~Durand and P.~Flajolet, ``Loglog counting of large cardinalities,'' in
  \emph{Proceedings of the 11th Annual European Symposium on Algorithms}, 2003.

\bibitem{Zhao2015c}
J.~Zhao, J.~C. Lui, D.~Towsley, P.~Wang, and X.~Guan, ``Tracking triadic
  cardinality distributions for burst detection in social activity streams,''
  in \emph{ACM Conferencec on Online Social Networks}, 2015.

\bibitem{Trees2001}
H.~L.~V. Trees, \emph{Detection, Estimation, and Modulation Theory, Part
  I}.\hskip 1em plus 0.5em minus 0.4em\relax Wiley-Interscience, 2001.

\bibitem{crlb}
``{Cram\'er-Rao} bound,''
  \url{https://en.wikipedia.org/wiki/Cram%C3%A9r%E2%80%93Rao_bound#Multivariate_case},
  Retrived Aug 2017.

\bibitem{Gorman1990}
J.~D. Gorman and A.~O. Hero, ``Lower bounds for parametric estimation with
  constraints,'' \emph{IEEE Transition on Information Theory}, vol.~26, no.~6,
  pp. 1285--1301, 1990.

\bibitem{Klimt2004}
B.~Klimt and Y.~Yang, ``The {Enron} corpus: A new dataset for email
  classification research,'' in \emph{Proceeding of the European Conference on
  Machine Learning and Principles and Practice of Knowledge Discovery in
  Databases}, 2004.

\bibitem{Backstrom2011}
L.~Backstrom, E.~Bakshy, J.~Kleinberg, T.~M. Lento, and I.~Rosenn, ``Center of
  attention: How {Facebook} users allocate attention across friends,'' in
  \emph{Proceedings of the 5th International AAAI Conference on Weblogs and
  Social Media}, 2011.

\bibitem{Mathioudakis2010}
M.~Mathioudakis, N.~Bansal, and N.~Koudas, ``Identifying, attributing and
  describing spatial bursts,'' in \emph{Proceedings of the VLDB Endowment},
  2010.

\bibitem{Zhu2003}
Y.~Zhu and D.~Shasha, ``Efficient elastic burst detection in data streams,'' in
  \emph{Proceedings of the 9th ACM SIGKDD International Conference on Knowledge
  Discovery and Data Mining}, 2003.

\bibitem{Barabasi2005}
A.-L. Barabasi, ``The origin of bursts and heavy tails in human dynamics,''
  \emph{Nature}, vol. 435, pp. 207--211, 2005.

\bibitem{Chierichetti2014}
F.~Chierichetti, J.~Kleinberg, R.~Kumar, M.~Mahdian, and S.~Pandey, ``Event
  detection via communication pattern analysis,'' in \emph{Proceedings of the
  8th International AAAI Conference on Weblogs and Social Media}, 2014.

\bibitem{Takahashi2011}
T.~Takahashi, R.~Tomioka, and K.~Yamanishi, ``Discovering emerging topics in
  social streams via link anomaly detection,'' in \emph{Proceedings of the IEEE
  International Conference on Data Mining}, 2011.

\bibitem{Sakaki2010}
T.~Sakaki, M.~Okazaki, and Y.~Matsuo, ``Earthquake shakes {Twitter} users:
  Real-time event detection by social sensors,'' in \emph{Proceedings of the
  19th International World Wide Web Conference}, 2010.

\bibitem{Manzoor2016}
E.~Manzoor, S.~M. Milajerdi, and L.~Akoglu, ``Fast memory-efficient anomaly
  detection in streaming heterogeneous graphs,'' in \emph{KDD}, San Francisco,
  California, USA, 2016.

\bibitem{Milo2002}
R.~Milo, S.~Shen-Orr, S.~Itzkovitz, N.~Kashtan, D.~Chklovskii, and U.~Alon,
  ``Network motifs: Simple building blocks of complex networks,''
  \emph{Science}, vol. 298, no. 5594, pp. 824--827, 2002.

\bibitem{Turkett2011}
W.~Turkett, E.~Fulp, C.~Lever, and J.~Edward~Allan, ``Graph mining of motif
  profiles for computer network activity inference,'' in \emph{Proceedings of
  the 7th Workshop on Mining and Learning with Graphs}, 2011.

\bibitem{Wang2014}
P.~Wang, J.~C. Lui, B.~Ribeiro, D.~Towsley, J.~Zhao, and X.~Guan, ``Efficiently
  estimating motif statistics of large networks,'' \emph{ACM Transactions on
  Knowledge Discovery from Data}, vol.~9, no.~2, pp. 1--27, 2014.

\bibitem{Budak2011}
C.~Budak, D.~Agrawal, and A.~E. Abbadi, ``Structural trend analysis for online
  social networks,'' in \emph{Proceedings of the VLDB Endowment}, 2011.

\bibitem{Pavan2013}
A.~Pavan, K.~Tangwongsan, S.~Tirthapura, and K.-L. Wu, ``Counting and sampling
  triangles from a graph stream,'' in \emph{Proceedings of the VLDB Endowment},
  2013.

\bibitem{Jha2013}
M.~Jha, C.~Seshadhri, and A.~Pinar, ``A space efficient streaming algorithm for
  triangle counting using the birthday paradox,'' in \emph{Proceedings of the
  19th ACM SIGKDD International Conference on Knowledge Discovery and Data
  Mining}, 2013.

\bibitem{Lim2015a}
Y.~Lim and U.~Kang, ``{MASCOT}: Memory-efficient and accurate sampling for
  counting local triangles in graph streams,'' in \emph{Proceedings of the 21st
  ACM SIGKDD International Conference on Knowledge Discovery and Data Mining},
  Sydney, Australia, 2015.

\bibitem{Stefani2016a}
L.~D. Stefani, A.~Epasto, M.~Riondato, and E.~Upfal, ``{TRIEST}: Counting local
  and global triangles in fully-dynamic streams with fixed memory size,'' in
  \emph{Proceedings of the 22nd ACM SIGKDD International Conference on
  Knowledge Discovery and Data Mining}, 2016.

\bibitem{Wu2016b}
B.~Wu, K.~Yi, and Z.~Li, ``Counting triangles in large graphs by random
  sampling,'' \emph{IEEE Transactions on Knowledge and Data Engineering},
  vol.~28, no.~8, pp. 2013--2026, 2016.

\bibitem{Seshadhri2013}
C.~Seshadhri, A.~Pinar, and T.~G. Kolda, ``Triadic measures on graphs: The
  power of wedge sampling,'' in \emph{Proceedings of the 13th SIAM
  International Conference on Data Mining}, 2013.

\bibitem{Becchetti2008}
L.~Becchetti, P.~Boldi, C.~Castillo, and A.~Gionis, ``Efficient semi-streaming
  algorithms for local triangle counting in massive graphs,'' in
  \emph{Proceedings of the 14th ACM SIGKDD International Conference on
  Knowledge Discovery and Data Mining}, 2008.

\end{thebibliography}

\end{document}